\documentclass[preprint,aps,amsmath,superscriptaddress,nofootinbib,tightenlines]{revtex4}
\usepackage{graphicx}
\usepackage{bm}
\usepackage{epsfig}
\usepackage{ulem}
\usepackage{color}
\definecolor{AliceBlue}{rgb}{0.94,0.97,1.00}
\definecolor{AntiqueWhite1}{rgb}{1.00,0.94,0.86}
\definecolor{AntiqueWhite2}{rgb}{0.93,0.87,0.80}
\definecolor{AntiqueWhite3}{rgb}{0.80,0.75,0.69}
\definecolor{AntiqueWhite4}{rgb}{0.55,0.51,0.47}
\definecolor{AntiqueWhite}{rgb}{0.98,0.92,0.84}
\definecolor{BlanchedAlmond}{rgb}{1.00,0.92,0.80}
\definecolor{BlueViolet}{rgb}{0.54,0.17,0.89}
\definecolor{CadetBlue1}{rgb}{0.60,0.96,1.00}
\definecolor{CadetBlue2}{rgb}{0.56,0.90,0.93}
\definecolor{CadetBlue3}{rgb}{0.48,0.77,0.80}
\definecolor{CadetBlue4}{rgb}{0.33,0.53,0.55}
\definecolor{CadetBlue}{rgb}{0.37,0.62,0.63}
\definecolor{CornflowerBlue}{rgb}{0.39,0.58,0.93}
\definecolor{DarkBlue}{rgb}{0.00,0.00,0.55}
\definecolor{DarkCyan}{rgb}{0.00,0.55,0.55}
\definecolor{DarkGoldenrod1}{rgb}{1.00,0.73,0.06}
\definecolor{DarkGoldenrod2}{rgb}{0.93,0.68,0.05}
\definecolor{DarkGoldenrod3}{rgb}{0.80,0.58,0.05}
\definecolor{DarkGoldenrod4}{rgb}{0.55,0.40,0.03}
\definecolor{DarkGoldenrod}{rgb}{0.72,0.53,0.04}
\definecolor{DarkGray}{rgb}{0.66,0.66,0.66}
\definecolor{DarkGreen}{rgb}{0.00,0.39,0.00}
\definecolor{DarkGrey}{rgb}{0.66,0.66,0.66}
\definecolor{DarkKhaki}{rgb}{0.74,0.72,0.42}
\definecolor{DarkMagenta}{rgb}{0.55,0.00,0.55}
\definecolor{DarkOliveGreen1}{rgb}{0.79,1.00,0.44}
\definecolor{DarkOliveGreen2}{rgb}{0.74,0.93,0.41}
\definecolor{DarkOliveGreen3}{rgb}{0.64,0.80,0.35}
\definecolor{DarkOliveGreen4}{rgb}{0.43,0.55,0.24}
\definecolor{DarkOliveGreen}{rgb}{0.33,0.42,0.18}
\definecolor{DarkOrange1}{rgb}{1.00,0.50,0.00}
\definecolor{DarkOrange2}{rgb}{0.93,0.46,0.00}
\definecolor{DarkOrange3}{rgb}{0.80,0.40,0.00}
\definecolor{DarkOrange4}{rgb}{0.55,0.27,0.00}
\definecolor{DarkOrange}{rgb}{1.00,0.55,0.00}
\definecolor{DarkOrchid1}{rgb}{0.75,0.24,1.00}
\definecolor{DarkOrchid2}{rgb}{0.70,0.23,0.93}
\definecolor{DarkOrchid3}{rgb}{0.60,0.20,0.80}
\definecolor{DarkOrchid4}{rgb}{0.41,0.13,0.55}
\definecolor{DarkOrchid}{rgb}{0.60,0.20,0.80}
\definecolor{DarkRed}{rgb}{0.55,0.00,0.00}
\definecolor{DarkSalmon}{rgb}{0.91,0.59,0.48}
\definecolor{DarkSeaGreen1}{rgb}{0.76,1.00,0.76}
\definecolor{DarkSeaGreen2}{rgb}{0.71,0.93,0.71}
\definecolor{DarkSeaGreen3}{rgb}{0.61,0.80,0.61}
\definecolor{DarkSeaGreen4}{rgb}{0.41,0.55,0.41}
\definecolor{DarkSeaGreen}{rgb}{0.56,0.74,0.56}
\definecolor{DarkSlateBlue}{rgb}{0.28,0.24,0.55}
\definecolor{DarkSlateGray1}{rgb}{0.59,1.00,1.00}
\definecolor{DarkSlateGray2}{rgb}{0.55,0.93,0.93}
\definecolor{DarkSlateGray3}{rgb}{0.47,0.80,0.80}
\definecolor{DarkSlateGray4}{rgb}{0.32,0.55,0.55}
\definecolor{DarkSlateGray}{rgb}{0.18,0.31,0.31}
\definecolor{DarkSlateGrey}{rgb}{0.18,0.31,0.31}
\definecolor{DarkTurquoise}{rgb}{0.00,0.81,0.82}
\definecolor{DarkViolet}{rgb}{0.58,0.00,0.83}
\definecolor{DeepPink1}{rgb}{1.00,0.08,0.58}
\definecolor{DeepPink2}{rgb}{0.93,0.07,0.54}
\definecolor{DeepPink3}{rgb}{0.80,0.06,0.46}
\definecolor{DeepPink4}{rgb}{0.55,0.04,0.31}
\definecolor{DeepPink}{rgb}{1.00,0.08,0.58}
\definecolor{DeepSkyBlue1}{rgb}{0.00,0.75,1.00}
\definecolor{DeepSkyBlue2}{rgb}{0.00,0.70,0.93}
\definecolor{DeepSkyBlue3}{rgb}{0.00,0.60,0.80}
\definecolor{DeepSkyBlue4}{rgb}{0.00,0.41,0.55}
\definecolor{DeepSkyBlue}{rgb}{0.00,0.75,1.00}
\definecolor{DimGray}{rgb}{0.41,0.41,0.41}
\definecolor{DimGrey}{rgb}{0.41,0.41,0.41}
\definecolor{DodgerBlue1}{rgb}{0.12,0.56,1.00}
\definecolor{DodgerBlue2}{rgb}{0.11,0.53,0.93}
\definecolor{DodgerBlue3}{rgb}{0.09,0.45,0.80}
\definecolor{DodgerBlue4}{rgb}{0.06,0.31,0.55}
\definecolor{DodgerBlue}{rgb}{0.12,0.56,1.00}
\definecolor{FloralWhite}{rgb}{1.00,0.98,0.94}
\definecolor{ForestGreen}{rgb}{0.13,0.55,0.13}
\definecolor{GhostWhite}{rgb}{0.97,0.97,1.00}
\definecolor{GreenYellow}{rgb}{0.68,1.00,0.18}
\definecolor{HotPink1}{rgb}{1.00,0.43,0.71}
\definecolor{HotPink2}{rgb}{0.93,0.42,0.65}
\definecolor{HotPink3}{rgb}{0.80,0.38,0.56}
\definecolor{HotPink4}{rgb}{0.55,0.23,0.38}
\definecolor{HotPink}{rgb}{1.00,0.41,0.71}
\definecolor{IndianRed1}{rgb}{1.00,0.42,0.42}
\definecolor{IndianRed2}{rgb}{0.93,0.39,0.39}
\definecolor{IndianRed3}{rgb}{0.80,0.33,0.33}
\definecolor{IndianRed4}{rgb}{0.55,0.23,0.23}
\definecolor{IndianRed}{rgb}{0.80,0.36,0.36}
\definecolor{LavenderBlush1}{rgb}{1.00,0.94,0.96}
\definecolor{LavenderBlush2}{rgb}{0.93,0.88,0.90}
\definecolor{LavenderBlush3}{rgb}{0.80,0.76,0.77}
\definecolor{LavenderBlush4}{rgb}{0.55,0.51,0.53}
\definecolor{LavenderBlush}{rgb}{1.00,0.94,0.96}
\definecolor{LawnGreen}{rgb}{0.49,0.99,0.00}
\definecolor{LemonChiffon1}{rgb}{1.00,0.98,0.80}
\definecolor{LemonChiffon2}{rgb}{0.93,0.91,0.75}
\definecolor{LemonChiffon3}{rgb}{0.80,0.79,0.65}
\definecolor{LemonChiffon4}{rgb}{0.55,0.54,0.44}
\definecolor{LemonChiffon}{rgb}{1.00,0.98,0.80}
\definecolor{LightBlue1}{rgb}{0.75,0.94,1.00}
\definecolor{LightBlue2}{rgb}{0.70,0.87,0.93}
\definecolor{LightBlue3}{rgb}{0.60,0.75,0.80}
\definecolor{LightBlue4}{rgb}{0.41,0.51,0.55}
\definecolor{LightBlue}{rgb}{0.68,0.85,0.90}
\definecolor{LightCoral}{rgb}{0.94,0.50,0.50}
\definecolor{LightCyan1}{rgb}{0.88,1.00,1.00}
\definecolor{LightCyan2}{rgb}{0.82,0.93,0.93}
\definecolor{LightCyan3}{rgb}{0.71,0.80,0.80}
\definecolor{LightCyan4}{rgb}{0.48,0.55,0.55}
\definecolor{LightCyan}{rgb}{0.88,1.00,1.00}
\definecolor{LightGoldenrod1}{rgb}{1.00,0.93,0.55}
\definecolor{LightGoldenrod2}{rgb}{0.93,0.86,0.51}
\definecolor{LightGoldenrod3}{rgb}{0.80,0.75,0.44}
\definecolor{LightGoldenrod4}{rgb}{0.55,0.51,0.30}
\definecolor{LightGoldenrodYellow}{rgb}{0.98,0.98,0.82}
\definecolor{LightGoldenrod}{rgb}{0.93,0.87,0.51}
\definecolor{LightGray}{rgb}{0.83,0.83,0.83}
\definecolor{LightGreen}{rgb}{0.56,0.93,0.56}
\definecolor{LightGrey}{rgb}{0.83,0.83,0.83}
\definecolor{LightPink1}{rgb}{1.00,0.68,0.73}
\definecolor{LightPink2}{rgb}{0.93,0.64,0.68}
\definecolor{LightPink3}{rgb}{0.80,0.55,0.58}
\definecolor{LightPink4}{rgb}{0.55,0.37,0.40}
\definecolor{LightPink}{rgb}{1.00,0.71,0.76}
\definecolor{LightSalmon1}{rgb}{1.00,0.63,0.48}
\definecolor{LightSalmon2}{rgb}{0.93,0.58,0.45}
\definecolor{LightSalmon3}{rgb}{0.80,0.51,0.38}
\definecolor{LightSalmon4}{rgb}{0.55,0.34,0.26}
\definecolor{LightSalmon}{rgb}{1.00,0.63,0.48}
\definecolor{LightSeaGreen}{rgb}{0.13,0.70,0.67}
\definecolor{LightSkyBlue1}{rgb}{0.69,0.89,1.00}
\definecolor{LightSkyBlue2}{rgb}{0.64,0.83,0.93}
\definecolor{LightSkyBlue3}{rgb}{0.55,0.71,0.80}
\definecolor{LightSkyBlue4}{rgb}{0.38,0.48,0.55}
\definecolor{LightSkyBlue}{rgb}{0.53,0.81,0.98}
\definecolor{LightSlateBlue}{rgb}{0.52,0.44,1.00}
\definecolor{LightSlateGray}{rgb}{0.47,0.53,0.60}
\definecolor{LightSlateGrey}{rgb}{0.47,0.53,0.60}
\definecolor{LightSteelBlue1}{rgb}{0.79,0.88,1.00}
\definecolor{LightSteelBlue2}{rgb}{0.74,0.82,0.93}
\definecolor{LightSteelBlue3}{rgb}{0.64,0.71,0.80}
\definecolor{LightSteelBlue4}{rgb}{0.43,0.48,0.55}
\definecolor{LightSteelBlue}{rgb}{0.69,0.77,0.87}
\definecolor{LightYellow1}{rgb}{1.00,1.00,0.88}
\definecolor{LightYellow2}{rgb}{0.93,0.93,0.82}
\definecolor{LightYellow3}{rgb}{0.80,0.80,0.71}
\definecolor{LightYellow4}{rgb}{0.55,0.55,0.48}
\definecolor{LightYellow}{rgb}{1.00,1.00,0.88}
\definecolor{LimeGreen}{rgb}{0.20,0.80,0.20}
\definecolor{MediumAquamarine}{rgb}{0.40,0.80,0.67}
\definecolor{MediumBlue}{rgb}{0.00,0.00,0.80}
\definecolor{MediumOrchid1}{rgb}{0.88,0.40,1.00}
\definecolor{MediumOrchid2}{rgb}{0.82,0.37,0.93}
\definecolor{MediumOrchid3}{rgb}{0.71,0.32,0.80}
\definecolor{MediumOrchid4}{rgb}{0.48,0.22,0.55}
\definecolor{MediumOrchid}{rgb}{0.73,0.33,0.83}
\definecolor{MediumPurple1}{rgb}{0.67,0.51,1.00}
\definecolor{MediumPurple2}{rgb}{0.62,0.47,0.93}
\definecolor{MediumPurple3}{rgb}{0.54,0.41,0.80}
\definecolor{MediumPurple4}{rgb}{0.36,0.28,0.55}
\definecolor{MediumPurple}{rgb}{0.58,0.44,0.86}
\definecolor{MediumSeaGreen}{rgb}{0.24,0.70,0.44}
\definecolor{MediumSlateBlue}{rgb}{0.48,0.41,0.93}
\definecolor{MediumSpringGreen}{rgb}{0.00,0.98,0.60}
\definecolor{MediumTurquoise}{rgb}{0.28,0.82,0.80}
\definecolor{MediumVioletRed}{rgb}{0.78,0.08,0.52}
\definecolor{MidnightBlue}{rgb}{0.10,0.10,0.44}
\definecolor{MintCream}{rgb}{0.96,1.00,0.98}
\definecolor{MistyRose1}{rgb}{1.00,0.89,0.88}
\definecolor{MistyRose2}{rgb}{0.93,0.84,0.82}
\definecolor{MistyRose3}{rgb}{0.80,0.72,0.71}
\definecolor{MistyRose4}{rgb}{0.55,0.49,0.48}
\definecolor{MistyRose}{rgb}{1.00,0.89,0.88}
\definecolor{NavajoWhite1}{rgb}{1.00,0.87,0.68}
\definecolor{NavajoWhite2}{rgb}{0.93,0.81,0.63}
\definecolor{NavajoWhite3}{rgb}{0.80,0.70,0.55}
\definecolor{NavajoWhite4}{rgb}{0.55,0.47,0.37}
\definecolor{NavajoWhite}{rgb}{1.00,0.87,0.68}
\definecolor{NavyBlue}{rgb}{0.00,0.00,0.50}
\definecolor{OldLace}{rgb}{0.99,0.96,0.90}
\definecolor{OliveDrab1}{rgb}{0.75,1.00,0.24}
\definecolor{OliveDrab2}{rgb}{0.70,0.93,0.23}
\definecolor{OliveDrab3}{rgb}{0.60,0.80,0.20}
\definecolor{OliveDrab4}{rgb}{0.41,0.55,0.13}
\definecolor{OliveDrab}{rgb}{0.42,0.56,0.14}
\definecolor{OrangeRed1}{rgb}{1.00,0.27,0.00}
\definecolor{OrangeRed2}{rgb}{0.93,0.25,0.00}
\definecolor{OrangeRed3}{rgb}{0.80,0.22,0.00}
\definecolor{OrangeRed4}{rgb}{0.55,0.15,0.00}
\definecolor{OrangeRed}{rgb}{1.00,0.27,0.00}
\definecolor{PaleGoldenrod}{rgb}{0.93,0.91,0.67}
\definecolor{PaleGreen1}{rgb}{0.60,1.00,0.60}
\definecolor{PaleGreen2}{rgb}{0.56,0.93,0.56}
\definecolor{PaleGreen3}{rgb}{0.49,0.80,0.49}
\definecolor{PaleGreen4}{rgb}{0.33,0.55,0.33}
\definecolor{PaleGreen}{rgb}{0.60,0.98,0.60}
\definecolor{PaleTurquoise1}{rgb}{0.73,1.00,1.00}
\definecolor{PaleTurquoise2}{rgb}{0.68,0.93,0.93}
\definecolor{PaleTurquoise3}{rgb}{0.59,0.80,0.80}
\definecolor{PaleTurquoise4}{rgb}{0.40,0.55,0.55}
\definecolor{PaleTurquoise}{rgb}{0.69,0.93,0.93}
\definecolor{PaleVioletRed1}{rgb}{1.00,0.51,0.67}
\definecolor{PaleVioletRed2}{rgb}{0.93,0.47,0.62}
\definecolor{PaleVioletRed3}{rgb}{0.80,0.41,0.54}
\definecolor{PaleVioletRed4}{rgb}{0.55,0.28,0.36}
\definecolor{PaleVioletRed}{rgb}{0.86,0.44,0.58}
\definecolor{PapayaWhip}{rgb}{1.00,0.94,0.84}
\definecolor{PeachPuff1}{rgb}{1.00,0.85,0.73}
\definecolor{PeachPuff2}{rgb}{0.93,0.80,0.68}
\definecolor{PeachPuff3}{rgb}{0.80,0.69,0.58}
\definecolor{PeachPuff4}{rgb}{0.55,0.47,0.40}
\definecolor{PeachPuff}{rgb}{1.00,0.85,0.73}
\definecolor{PowderBlue}{rgb}{0.69,0.88,0.90}
\definecolor{RosyBrown1}{rgb}{1.00,0.76,0.76}
\definecolor{RosyBrown2}{rgb}{0.93,0.71,0.71}
\definecolor{RosyBrown3}{rgb}{0.80,0.61,0.61}
\definecolor{RosyBrown4}{rgb}{0.55,0.41,0.41}
\definecolor{RosyBrown}{rgb}{0.74,0.56,0.56}
\definecolor{RoyalBlue1}{rgb}{0.28,0.46,1.00}
\definecolor{RoyalBlue2}{rgb}{0.26,0.43,0.93}
\definecolor{RoyalBlue3}{rgb}{0.23,0.37,0.80}
\definecolor{RoyalBlue4}{rgb}{0.15,0.25,0.55}
\definecolor{RoyalBlue}{rgb}{0.25,0.41,0.88}
\definecolor{SaddleBrown}{rgb}{0.55,0.27,0.07}
\definecolor{SandyBrown}{rgb}{0.96,0.64,0.38}
\definecolor{SeaGreen1}{rgb}{0.33,1.00,0.62}
\definecolor{SeaGreen2}{rgb}{0.31,0.93,0.58}
\definecolor{SeaGreen3}{rgb}{0.26,0.80,0.50}
\definecolor{SeaGreen4}{rgb}{0.18,0.55,0.34}
\definecolor{SeaGreen}{rgb}{0.18,0.55,0.34}
\definecolor{SkyBlue1}{rgb}{0.53,0.81,1.00}
\definecolor{SkyBlue2}{rgb}{0.49,0.75,0.93}
\definecolor{SkyBlue3}{rgb}{0.42,0.65,0.80}
\definecolor{SkyBlue4}{rgb}{0.29,0.44,0.55}
\definecolor{SkyBlue}{rgb}{0.53,0.81,0.92}
\definecolor{SlateBlue1}{rgb}{0.51,0.44,1.00}
\definecolor{SlateBlue2}{rgb}{0.48,0.40,0.93}
\definecolor{SlateBlue3}{rgb}{0.41,0.35,0.80}
\definecolor{SlateBlue4}{rgb}{0.28,0.24,0.55}
\definecolor{SlateBlue}{rgb}{0.42,0.35,0.80}
\definecolor{SlateGray1}{rgb}{0.78,0.89,1.00}
\definecolor{SlateGray2}{rgb}{0.73,0.83,0.93}
\definecolor{SlateGray3}{rgb}{0.62,0.71,0.80}
\definecolor{SlateGray4}{rgb}{0.42,0.48,0.55}
\definecolor{SlateGray}{rgb}{0.44,0.50,0.56}
\definecolor{SlateGrey}{rgb}{0.44,0.50,0.56}
\definecolor{SpringGreen1}{rgb}{0.00,1.00,0.50}
\definecolor{SpringGreen2}{rgb}{0.00,0.93,0.46}
\definecolor{SpringGreen3}{rgb}{0.00,0.80,0.40}
\definecolor{SpringGreen4}{rgb}{0.00,0.55,0.27}
\definecolor{SpringGreen}{rgb}{0.00,1.00,0.50}
\definecolor{SteelBlue1}{rgb}{0.39,0.72,1.00}
\definecolor{SteelBlue2}{rgb}{0.36,0.67,0.93}
\definecolor{SteelBlue3}{rgb}{0.31,0.58,0.80}
\definecolor{SteelBlue4}{rgb}{0.21,0.39,0.55}
\definecolor{SteelBlue}{rgb}{0.27,0.51,0.71}
\definecolor{VioletRed1}{rgb}{1.00,0.24,0.59}
\definecolor{VioletRed2}{rgb}{0.93,0.23,0.55}
\definecolor{VioletRed3}{rgb}{0.80,0.20,0.47}
\definecolor{VioletRed4}{rgb}{0.55,0.13,0.32}
\definecolor{VioletRed}{rgb}{0.82,0.13,0.56}
\definecolor{WhiteSmoke}{rgb}{0.96,0.96,0.96}
\definecolor{YellowGreen}{rgb}{0.60,0.80,0.20}
\definecolor{aliceblue}{rgb}{0.94,0.97,1.00}
\definecolor{antiquewhite}{rgb}{0.98,0.92,0.84}
\definecolor{aquamarine1}{rgb}{0.50,1.00,0.83}
\definecolor{aquamarine2}{rgb}{0.46,0.93,0.78}
\definecolor{aquamarine3}{rgb}{0.40,0.80,0.67}
\definecolor{aquamarine4}{rgb}{0.27,0.55,0.45}
\definecolor{aquamarine}{rgb}{0.50,1.00,0.83}
\definecolor{azure1}{rgb}{0.94,1.00,1.00}
\definecolor{azure2}{rgb}{0.88,0.93,0.93}
\definecolor{azure3}{rgb}{0.76,0.80,0.80}
\definecolor{azure4}{rgb}{0.51,0.55,0.55}
\definecolor{azure}{rgb}{0.94,1.00,1.00}
\definecolor{beige}{rgb}{0.96,0.96,0.86}
\definecolor{bisque1}{rgb}{1.00,0.89,0.77}
\definecolor{bisque2}{rgb}{0.93,0.84,0.72}
\definecolor{bisque3}{rgb}{0.80,0.72,0.62}
\definecolor{bisque4}{rgb}{0.55,0.49,0.42}
\definecolor{bisque}{rgb}{1.00,0.89,0.77}
\definecolor{black}{rgb}{0.00,0.00,0.00}
\definecolor{blanchedalmond}{rgb}{1.00,0.92,0.80}
\definecolor{blue1}{rgb}{0.00,0.00,1.00}
\definecolor{blue2}{rgb}{0.00,0.00,0.93}
\definecolor{blue3}{rgb}{0.00,0.00,0.80}
\definecolor{blue4}{rgb}{0.00,0.00,0.55}
\definecolor{blueviolet}{rgb}{0.54,0.17,0.89}
\definecolor{blue}{rgb}{0.00,0.00,1.00}
\definecolor{brown1}{rgb}{1.00,0.25,0.25}
\definecolor{brown2}{rgb}{0.93,0.23,0.23}
\definecolor{brown3}{rgb}{0.80,0.20,0.20}
\definecolor{brown4}{rgb}{0.55,0.14,0.14}
\definecolor{brown}{rgb}{0.65,0.16,0.16}
\definecolor{burlywood1}{rgb}{1.00,0.83,0.61}
\definecolor{burlywood2}{rgb}{0.93,0.77,0.57}
\definecolor{burlywood3}{rgb}{0.80,0.67,0.49}
\definecolor{burlywood4}{rgb}{0.55,0.45,0.33}
\definecolor{burlywood}{rgb}{0.87,0.72,0.53}
\definecolor{cadetblue}{rgb}{0.37,0.62,0.63}
\definecolor{chartreuse1}{rgb}{0.50,1.00,0.00}
\definecolor{chartreuse2}{rgb}{0.46,0.93,0.00}
\definecolor{chartreuse3}{rgb}{0.40,0.80,0.00}
\definecolor{chartreuse4}{rgb}{0.27,0.55,0.00}
\definecolor{chartreuse}{rgb}{0.50,1.00,0.00}
\definecolor{chocolate1}{rgb}{1.00,0.50,0.14}
\definecolor{chocolate2}{rgb}{0.93,0.46,0.13}
\definecolor{chocolate3}{rgb}{0.80,0.40,0.11}
\definecolor{chocolate4}{rgb}{0.55,0.27,0.07}
\definecolor{chocolate}{rgb}{0.82,0.41,0.12}
\definecolor{coral1}{rgb}{1.00,0.45,0.34}
\definecolor{coral2}{rgb}{0.93,0.42,0.31}
\definecolor{coral3}{rgb}{0.80,0.36,0.27}
\definecolor{coral4}{rgb}{0.55,0.24,0.18}
\definecolor{coral}{rgb}{1.00,0.50,0.31}
\definecolor{cornflowerblue}{rgb}{0.39,0.58,0.93}
\definecolor{cornsilk1}{rgb}{1.00,0.97,0.86}
\definecolor{cornsilk2}{rgb}{0.93,0.91,0.80}
\definecolor{cornsilk3}{rgb}{0.80,0.78,0.69}
\definecolor{cornsilk4}{rgb}{0.55,0.53,0.47}
\definecolor{cornsilk}{rgb}{1.00,0.97,0.86}
\definecolor{cyan1}{rgb}{0.00,1.00,1.00}
\definecolor{cyan2}{rgb}{0.00,0.93,0.93}
\definecolor{cyan3}{rgb}{0.00,0.80,0.80}
\definecolor{cyan4}{rgb}{0.00,0.55,0.55}
\definecolor{cyan}{rgb}{0.00,1.00,1.00}
\definecolor{darkblue}{rgb}{0.00,0.00,0.55}
\definecolor{darkcyan}{rgb}{0.00,0.55,0.55}
\definecolor{darkgoldenrod}{rgb}{0.72,0.53,0.04}
\definecolor{darkgray}{rgb}{0.66,0.66,0.66}
\definecolor{darkgreen}{rgb}{0.00,0.39,0.00}
\definecolor{darkgrey}{rgb}{0.66,0.66,0.66}
\definecolor{darkkhaki}{rgb}{0.74,0.72,0.42}
\definecolor{darkmagenta}{rgb}{0.55,0.00,0.55}
\definecolor{darkolive}{rgb}{0.33,0.42,0.18}
\definecolor{darkorange}{rgb}{1.00,0.55,0.00}
\definecolor{darkorchid}{rgb}{0.60,0.20,0.80}
\definecolor{darkred}{rgb}{0.55,0.00,0.00}
\definecolor{darksalmon}{rgb}{0.91,0.59,0.48}
\definecolor{darksea}{rgb}{0.56,0.74,0.56}
\definecolor{darkslate}{rgb}{0.18,0.31,0.31}
\definecolor{darkslate}{rgb}{0.18,0.31,0.31}
\definecolor{darkslate}{rgb}{0.28,0.24,0.55}
\definecolor{darkturquoise}{rgb}{0.00,0.81,0.82}
\definecolor{darkviolet}{rgb}{0.58,0.00,0.83}
\definecolor{deeppink}{rgb}{1.00,0.08,0.58}
\definecolor{deepsky}{rgb}{0.00,0.75,1.00}
\definecolor{dimgray}{rgb}{0.41,0.41,0.41}
\definecolor{dimgrey}{rgb}{0.41,0.41,0.41}
\definecolor{dodgerblue}{rgb}{0.12,0.56,1.00}
\definecolor{firebrick1}{rgb}{1.00,0.19,0.19}
\definecolor{firebrick2}{rgb}{0.93,0.17,0.17}
\definecolor{firebrick3}{rgb}{0.80,0.15,0.15}
\definecolor{firebrick4}{rgb}{0.55,0.10,0.10}
\definecolor{firebrick}{rgb}{0.70,0.13,0.13}
\definecolor{floralwhite}{rgb}{1.00,0.98,0.94}
\definecolor{forestgreen}{rgb}{0.13,0.55,0.13}
\definecolor{gainsboro}{rgb}{0.86,0.86,0.86}
\definecolor{ghostwhite}{rgb}{0.97,0.97,1.00}
\definecolor{gold1}{rgb}{1.00,0.84,0.00}
\definecolor{gold2}{rgb}{0.93,0.79,0.00}
\definecolor{gold3}{rgb}{0.80,0.68,0.00}
\definecolor{gold4}{rgb}{0.55,0.46,0.00}
\definecolor{goldenrod1}{rgb}{1.00,0.76,0.15}
\definecolor{goldenrod2}{rgb}{0.93,0.71,0.13}
\definecolor{goldenrod3}{rgb}{0.80,0.61,0.11}
\definecolor{goldenrod4}{rgb}{0.55,0.41,0.08}
\definecolor{goldenrod}{rgb}{0.85,0.65,0.13}
\definecolor{gold}{rgb}{1.00,0.84,0.00}
\definecolor{gray0}{rgb}{0.00,0.00,0.00}
\definecolor{gray100}{rgb}{1.00,1.00,1.00}
\definecolor{gray10}{rgb}{0.10,0.10,0.10}
\definecolor{gray11}{rgb}{0.11,0.11,0.11}
\definecolor{gray12}{rgb}{0.12,0.12,0.12}
\definecolor{gray13}{rgb}{0.13,0.13,0.13}
\definecolor{gray14}{rgb}{0.14,0.14,0.14}
\definecolor{gray15}{rgb}{0.15,0.15,0.15}
\definecolor{gray16}{rgb}{0.16,0.16,0.16}
\definecolor{gray17}{rgb}{0.17,0.17,0.17}
\definecolor{gray18}{rgb}{0.18,0.18,0.18}
\definecolor{gray19}{rgb}{0.19,0.19,0.19}
\definecolor{gray1}{rgb}{0.01,0.01,0.01}
\definecolor{gray20}{rgb}{0.20,0.20,0.20}
\definecolor{gray21}{rgb}{0.21,0.21,0.21}
\definecolor{gray22}{rgb}{0.22,0.22,0.22}
\definecolor{gray23}{rgb}{0.23,0.23,0.23}
\definecolor{gray24}{rgb}{0.24,0.24,0.24}
\definecolor{gray25}{rgb}{0.25,0.25,0.25}
\definecolor{gray26}{rgb}{0.26,0.26,0.26}
\definecolor{gray27}{rgb}{0.27,0.27,0.27}
\definecolor{gray28}{rgb}{0.28,0.28,0.28}
\definecolor{gray29}{rgb}{0.29,0.29,0.29}
\definecolor{gray2}{rgb}{0.02,0.02,0.02}
\definecolor{gray30}{rgb}{0.30,0.30,0.30}
\definecolor{gray31}{rgb}{0.31,0.31,0.31}
\definecolor{gray32}{rgb}{0.32,0.32,0.32}
\definecolor{gray33}{rgb}{0.33,0.33,0.33}
\definecolor{gray34}{rgb}{0.34,0.34,0.34}
\definecolor{gray35}{rgb}{0.35,0.35,0.35}
\definecolor{gray36}{rgb}{0.36,0.36,0.36}
\definecolor{gray37}{rgb}{0.37,0.37,0.37}
\definecolor{gray38}{rgb}{0.38,0.38,0.38}
\definecolor{gray39}{rgb}{0.39,0.39,0.39}
\definecolor{gray3}{rgb}{0.03,0.03,0.03}
\definecolor{gray40}{rgb}{0.40,0.40,0.40}
\definecolor{gray41}{rgb}{0.41,0.41,0.41}
\definecolor{gray42}{rgb}{0.42,0.42,0.42}
\definecolor{gray43}{rgb}{0.43,0.43,0.43}
\definecolor{gray44}{rgb}{0.44,0.44,0.44}
\definecolor{gray45}{rgb}{0.45,0.45,0.45}
\definecolor{gray46}{rgb}{0.46,0.46,0.46}
\definecolor{gray47}{rgb}{0.47,0.47,0.47}
\definecolor{gray48}{rgb}{0.48,0.48,0.48}
\definecolor{gray49}{rgb}{0.49,0.49,0.49}
\definecolor{gray4}{rgb}{0.04,0.04,0.04}
\definecolor{gray50}{rgb}{0.50,0.50,0.50}
\definecolor{gray51}{rgb}{0.51,0.51,0.51}
\definecolor{gray52}{rgb}{0.52,0.52,0.52}
\definecolor{gray53}{rgb}{0.53,0.53,0.53}
\definecolor{gray54}{rgb}{0.54,0.54,0.54}
\definecolor{gray55}{rgb}{0.55,0.55,0.55}
\definecolor{gray56}{rgb}{0.56,0.56,0.56}
\definecolor{gray57}{rgb}{0.57,0.57,0.57}
\definecolor{gray58}{rgb}{0.58,0.58,0.58}
\definecolor{gray59}{rgb}{0.59,0.59,0.59}
\definecolor{gray5}{rgb}{0.05,0.05,0.05}
\definecolor{gray60}{rgb}{0.60,0.60,0.60}
\definecolor{gray61}{rgb}{0.61,0.61,0.61}
\definecolor{gray62}{rgb}{0.62,0.62,0.62}
\definecolor{gray63}{rgb}{0.63,0.63,0.63}
\definecolor{gray64}{rgb}{0.64,0.64,0.64}
\definecolor{gray65}{rgb}{0.65,0.65,0.65}
\definecolor{gray66}{rgb}{0.66,0.66,0.66}
\definecolor{gray67}{rgb}{0.67,0.67,0.67}
\definecolor{gray68}{rgb}{0.68,0.68,0.68}
\definecolor{gray69}{rgb}{0.69,0.69,0.69}
\definecolor{gray6}{rgb}{0.06,0.06,0.06}
\definecolor{gray70}{rgb}{0.70,0.70,0.70}
\definecolor{gray71}{rgb}{0.71,0.71,0.71}
\definecolor{gray72}{rgb}{0.72,0.72,0.72}
\definecolor{gray73}{rgb}{0.73,0.73,0.73}
\definecolor{gray74}{rgb}{0.74,0.74,0.74}
\definecolor{gray75}{rgb}{0.75,0.75,0.75}
\definecolor{gray76}{rgb}{0.76,0.76,0.76}
\definecolor{gray77}{rgb}{0.77,0.77,0.77}
\definecolor{gray78}{rgb}{0.78,0.78,0.78}
\definecolor{gray79}{rgb}{0.79,0.79,0.79}
\definecolor{gray7}{rgb}{0.07,0.07,0.07}
\definecolor{gray80}{rgb}{0.80,0.80,0.80}
\definecolor{gray81}{rgb}{0.81,0.81,0.81}
\definecolor{gray82}{rgb}{0.82,0.82,0.82}
\definecolor{gray83}{rgb}{0.83,0.83,0.83}
\definecolor{gray84}{rgb}{0.84,0.84,0.84}
\definecolor{gray85}{rgb}{0.85,0.85,0.85}
\definecolor{gray86}{rgb}{0.86,0.86,0.86}
\definecolor{gray87}{rgb}{0.87,0.87,0.87}
\definecolor{gray88}{rgb}{0.88,0.88,0.88}
\definecolor{gray89}{rgb}{0.89,0.89,0.89}
\definecolor{gray8}{rgb}{0.08,0.08,0.08}
\definecolor{gray90}{rgb}{0.90,0.90,0.90}
\definecolor{gray91}{rgb}{0.91,0.91,0.91}
\definecolor{gray92}{rgb}{0.92,0.92,0.92}
\definecolor{gray93}{rgb}{0.93,0.93,0.93}
\definecolor{gray94}{rgb}{0.94,0.94,0.94}
\definecolor{gray95}{rgb}{0.95,0.95,0.95}
\definecolor{gray96}{rgb}{0.96,0.96,0.96}
\definecolor{gray97}{rgb}{0.97,0.97,0.97}
\definecolor{gray98}{rgb}{0.98,0.98,0.98}
\definecolor{gray99}{rgb}{0.99,0.99,0.99}
\definecolor{gray9}{rgb}{0.09,0.09,0.09}
\definecolor{gray}{rgb}{0.75,0.75,0.75}
\definecolor{green1}{rgb}{0.00,1.00,0.00}
\definecolor{green2}{rgb}{0.00,0.93,0.00}
\definecolor{green3}{rgb}{0.00,0.80,0.00}
\definecolor{green4}{rgb}{0.00,0.55,0.00}
\definecolor{greenyellow}{rgb}{0.68,1.00,0.18}
\definecolor{green}{rgb}{0.00,1.00,0.00}
\definecolor{grey0}{rgb}{0.00,0.00,0.00}
\definecolor{grey100}{rgb}{1.00,1.00,1.00}
\definecolor{grey10}{rgb}{0.10,0.10,0.10}
\definecolor{grey11}{rgb}{0.11,0.11,0.11}
\definecolor{grey12}{rgb}{0.12,0.12,0.12}
\definecolor{grey13}{rgb}{0.13,0.13,0.13}
\definecolor{grey14}{rgb}{0.14,0.14,0.14}
\definecolor{grey15}{rgb}{0.15,0.15,0.15}
\definecolor{grey16}{rgb}{0.16,0.16,0.16}
\definecolor{grey17}{rgb}{0.17,0.17,0.17}
\definecolor{grey18}{rgb}{0.18,0.18,0.18}
\definecolor{grey19}{rgb}{0.19,0.19,0.19}
\definecolor{grey1}{rgb}{0.01,0.01,0.01}
\definecolor{grey20}{rgb}{0.20,0.20,0.20}
\definecolor{grey21}{rgb}{0.21,0.21,0.21}
\definecolor{grey22}{rgb}{0.22,0.22,0.22}
\definecolor{grey23}{rgb}{0.23,0.23,0.23}
\definecolor{grey24}{rgb}{0.24,0.24,0.24}
\definecolor{grey25}{rgb}{0.25,0.25,0.25}
\definecolor{grey26}{rgb}{0.26,0.26,0.26}
\definecolor{grey27}{rgb}{0.27,0.27,0.27}
\definecolor{grey28}{rgb}{0.28,0.28,0.28}
\definecolor{grey29}{rgb}{0.29,0.29,0.29}
\definecolor{grey2}{rgb}{0.02,0.02,0.02}
\definecolor{grey30}{rgb}{0.30,0.30,0.30}
\definecolor{grey31}{rgb}{0.31,0.31,0.31}
\definecolor{grey32}{rgb}{0.32,0.32,0.32}
\definecolor{grey33}{rgb}{0.33,0.33,0.33}
\definecolor{grey34}{rgb}{0.34,0.34,0.34}
\definecolor{grey35}{rgb}{0.35,0.35,0.35}
\definecolor{grey36}{rgb}{0.36,0.36,0.36}
\definecolor{grey37}{rgb}{0.37,0.37,0.37}
\definecolor{grey38}{rgb}{0.38,0.38,0.38}
\definecolor{grey39}{rgb}{0.39,0.39,0.39}
\definecolor{grey3}{rgb}{0.03,0.03,0.03}
\definecolor{grey40}{rgb}{0.40,0.40,0.40}
\definecolor{grey41}{rgb}{0.41,0.41,0.41}
\definecolor{grey42}{rgb}{0.42,0.42,0.42}
\definecolor{grey43}{rgb}{0.43,0.43,0.43}
\definecolor{grey44}{rgb}{0.44,0.44,0.44}
\definecolor{grey45}{rgb}{0.45,0.45,0.45}
\definecolor{grey46}{rgb}{0.46,0.46,0.46}
\definecolor{grey47}{rgb}{0.47,0.47,0.47}
\definecolor{grey48}{rgb}{0.48,0.48,0.48}
\definecolor{grey49}{rgb}{0.49,0.49,0.49}
\definecolor{grey4}{rgb}{0.04,0.04,0.04}
\definecolor{grey50}{rgb}{0.50,0.50,0.50}
\definecolor{grey51}{rgb}{0.51,0.51,0.51}
\definecolor{grey52}{rgb}{0.52,0.52,0.52}
\definecolor{grey53}{rgb}{0.53,0.53,0.53}
\definecolor{grey54}{rgb}{0.54,0.54,0.54}
\definecolor{grey55}{rgb}{0.55,0.55,0.55}
\definecolor{grey56}{rgb}{0.56,0.56,0.56}
\definecolor{grey57}{rgb}{0.57,0.57,0.57}
\definecolor{grey58}{rgb}{0.58,0.58,0.58}
\definecolor{grey59}{rgb}{0.59,0.59,0.59}
\definecolor{grey5}{rgb}{0.05,0.05,0.05}
\definecolor{grey60}{rgb}{0.60,0.60,0.60}
\definecolor{grey61}{rgb}{0.61,0.61,0.61}
\definecolor{grey62}{rgb}{0.62,0.62,0.62}
\definecolor{grey63}{rgb}{0.63,0.63,0.63}
\definecolor{grey64}{rgb}{0.64,0.64,0.64}
\definecolor{grey65}{rgb}{0.65,0.65,0.65}
\definecolor{grey66}{rgb}{0.66,0.66,0.66}
\definecolor{grey67}{rgb}{0.67,0.67,0.67}
\definecolor{grey68}{rgb}{0.68,0.68,0.68}
\definecolor{grey69}{rgb}{0.69,0.69,0.69}
\definecolor{grey6}{rgb}{0.06,0.06,0.06}
\definecolor{grey70}{rgb}{0.70,0.70,0.70}
\definecolor{grey71}{rgb}{0.71,0.71,0.71}
\definecolor{grey72}{rgb}{0.72,0.72,0.72}
\definecolor{grey73}{rgb}{0.73,0.73,0.73}
\definecolor{grey74}{rgb}{0.74,0.74,0.74}
\definecolor{grey75}{rgb}{0.75,0.75,0.75}
\definecolor{grey76}{rgb}{0.76,0.76,0.76}
\definecolor{grey77}{rgb}{0.77,0.77,0.77}
\definecolor{grey78}{rgb}{0.78,0.78,0.78}
\definecolor{grey79}{rgb}{0.79,0.79,0.79}
\definecolor{grey7}{rgb}{0.07,0.07,0.07}
\definecolor{grey80}{rgb}{0.80,0.80,0.80}
\definecolor{grey81}{rgb}{0.81,0.81,0.81}
\definecolor{grey82}{rgb}{0.82,0.82,0.82}
\definecolor{grey83}{rgb}{0.83,0.83,0.83}
\definecolor{grey84}{rgb}{0.84,0.84,0.84}
\definecolor{grey85}{rgb}{0.85,0.85,0.85}
\definecolor{grey86}{rgb}{0.86,0.86,0.86}
\definecolor{grey87}{rgb}{0.87,0.87,0.87}
\definecolor{grey88}{rgb}{0.88,0.88,0.88}
\definecolor{grey89}{rgb}{0.89,0.89,0.89}
\definecolor{grey8}{rgb}{0.08,0.08,0.08}
\definecolor{grey90}{rgb}{0.90,0.90,0.90}
\definecolor{grey91}{rgb}{0.91,0.91,0.91}
\definecolor{grey92}{rgb}{0.92,0.92,0.92}
\definecolor{grey93}{rgb}{0.93,0.93,0.93}
\definecolor{grey94}{rgb}{0.94,0.94,0.94}
\definecolor{grey95}{rgb}{0.95,0.95,0.95}
\definecolor{grey96}{rgb}{0.96,0.96,0.96}
\definecolor{grey97}{rgb}{0.97,0.97,0.97}
\definecolor{grey98}{rgb}{0.98,0.98,0.98}
\definecolor{grey99}{rgb}{0.99,0.99,0.99}
\definecolor{grey9}{rgb}{0.09,0.09,0.09}
\definecolor{grey}{rgb}{0.75,0.75,0.75}
\definecolor{honeydew1}{rgb}{0.94,1.00,0.94}
\definecolor{honeydew2}{rgb}{0.88,0.93,0.88}
\definecolor{honeydew3}{rgb}{0.76,0.80,0.76}
\definecolor{honeydew4}{rgb}{0.51,0.55,0.51}
\definecolor{honeydew}{rgb}{0.94,1.00,0.94}
\definecolor{hotpink}{rgb}{1.00,0.41,0.71}
\definecolor{indianred}{rgb}{0.80,0.36,0.36}
\definecolor{ivory1}{rgb}{1.00,1.00,0.94}
\definecolor{ivory2}{rgb}{0.93,0.93,0.88}
\definecolor{ivory3}{rgb}{0.80,0.80,0.76}
\definecolor{ivory4}{rgb}{0.55,0.55,0.51}
\definecolor{ivory}{rgb}{1.00,1.00,0.94}
\definecolor{khaki1}{rgb}{1.00,0.96,0.56}
\definecolor{khaki2}{rgb}{0.93,0.90,0.52}
\definecolor{khaki3}{rgb}{0.80,0.78,0.45}
\definecolor{khaki4}{rgb}{0.55,0.53,0.31}
\definecolor{khaki}{rgb}{0.94,0.90,0.55}
\definecolor{lavenderblush}{rgb}{1.00,0.94,0.96}
\definecolor{lavender}{rgb}{0.90,0.90,0.98}
\definecolor{lawngreen}{rgb}{0.49,0.99,0.00}
\definecolor{lemonchiffon}{rgb}{1.00,0.98,0.80}
\definecolor{lightblue}{rgb}{0.68,0.85,0.90}
\definecolor{lightcoral}{rgb}{0.94,0.50,0.50}
\definecolor{lightcyan}{rgb}{0.88,1.00,1.00}
\definecolor{lightgoldenrod}{rgb}{0.93,0.87,0.51}
\definecolor{lightgoldenrod}{rgb}{0.98,0.98,0.82}
\definecolor{lightgray}{rgb}{0.83,0.83,0.83}
\definecolor{lightgreen}{rgb}{0.56,0.93,0.56}
\definecolor{lightgrey}{rgb}{0.83,0.83,0.83}
\definecolor{lightpink}{rgb}{1.00,0.71,0.76}
\definecolor{lightsalmon}{rgb}{1.00,0.63,0.48}
\definecolor{lightsea}{rgb}{0.13,0.70,0.67}
\definecolor{lightsky}{rgb}{0.53,0.81,0.98}
\definecolor{lightslate}{rgb}{0.47,0.53,0.60}
\definecolor{lightslate}{rgb}{0.47,0.53,0.60}
\definecolor{lightslate}{rgb}{0.52,0.44,1.00}
\definecolor{lightsteel}{rgb}{0.69,0.77,0.87}
\definecolor{lightyellow}{rgb}{1.00,1.00,0.88}
\definecolor{limegreen}{rgb}{0.20,0.80,0.20}
\definecolor{linen}{rgb}{0.98,0.94,0.90}
\definecolor{magenta1}{rgb}{1.00,0.00,1.00}
\definecolor{magenta2}{rgb}{0.93,0.00,0.93}
\definecolor{magenta3}{rgb}{0.80,0.00,0.80}
\definecolor{magenta4}{rgb}{0.55,0.00,0.55}
\definecolor{magenta}{rgb}{1.00,0.00,1.00}
\definecolor{maroon1}{rgb}{1.00,0.20,0.70}
\definecolor{maroon2}{rgb}{0.93,0.19,0.65}
\definecolor{maroon3}{rgb}{0.80,0.16,0.56}
\definecolor{maroon4}{rgb}{0.55,0.11,0.38}
\definecolor{maroon}{rgb}{0.69,0.19,0.38}
\definecolor{mediumaquamarine}{rgb}{0.40,0.80,0.67}
\definecolor{mediumblue}{rgb}{0.00,0.00,0.80}
\definecolor{mediumorchid}{rgb}{0.73,0.33,0.83}
\definecolor{mediumpurple}{rgb}{0.58,0.44,0.86}
\definecolor{mediumsea}{rgb}{0.24,0.70,0.44}
\definecolor{mediumslate}{rgb}{0.48,0.41,0.93}
\definecolor{mediumspring}{rgb}{0.00,0.98,0.60}
\definecolor{mediumturquoise}{rgb}{0.28,0.82,0.80}
\definecolor{mediumviolet}{rgb}{0.78,0.08,0.52}
\definecolor{midnightblue}{rgb}{0.10,0.10,0.44}
\definecolor{mintcream}{rgb}{0.96,1.00,0.98}
\definecolor{mistyrose}{rgb}{1.00,0.89,0.88}
\definecolor{moccasin}{rgb}{1.00,0.89,0.71}
\definecolor{navajowhite}{rgb}{1.00,0.87,0.68}
\definecolor{navyblue}{rgb}{0.00,0.00,0.50}
\definecolor{navy}{rgb}{0.00,0.00,0.50}
\definecolor{oldlace}{rgb}{0.99,0.96,0.90}
\definecolor{olivedrab}{rgb}{0.42,0.56,0.14}
\definecolor{orange1}{rgb}{1.00,0.65,0.00}
\definecolor{orange2}{rgb}{0.93,0.60,0.00}
\definecolor{orange3}{rgb}{0.80,0.52,0.00}
\definecolor{orange4}{rgb}{0.55,0.35,0.00}
\definecolor{orangered}{rgb}{1.00,0.27,0.00}
\definecolor{orange}{rgb}{1.00,0.65,0.00}
\definecolor{orchid1}{rgb}{1.00,0.51,0.98}
\definecolor{orchid2}{rgb}{0.93,0.48,0.91}
\definecolor{orchid3}{rgb}{0.80,0.41,0.79}
\definecolor{orchid4}{rgb}{0.55,0.28,0.54}
\definecolor{orchid}{rgb}{0.85,0.44,0.84}
\definecolor{palegoldenrod}{rgb}{0.93,0.91,0.67}
\definecolor{palegreen}{rgb}{0.60,0.98,0.60}
\definecolor{paleturquoise}{rgb}{0.69,0.93,0.93}
\definecolor{paleviolet}{rgb}{0.86,0.44,0.58}
\definecolor{papayawhip}{rgb}{1.00,0.94,0.84}
\definecolor{peachpuff}{rgb}{1.00,0.85,0.73}
\definecolor{peru}{rgb}{0.80,0.52,0.25}
\definecolor{pink1}{rgb}{1.00,0.71,0.77}
\definecolor{pink2}{rgb}{0.93,0.66,0.72}
\definecolor{pink3}{rgb}{0.80,0.57,0.62}
\definecolor{pink4}{rgb}{0.55,0.39,0.42}
\definecolor{pink}{rgb}{1.00,0.75,0.80}
\definecolor{plum1}{rgb}{1.00,0.73,1.00}
\definecolor{plum2}{rgb}{0.93,0.68,0.93}
\definecolor{plum3}{rgb}{0.80,0.59,0.80}
\definecolor{plum4}{rgb}{0.55,0.40,0.55}
\definecolor{plum}{rgb}{0.87,0.63,0.87}
\definecolor{powderblue}{rgb}{0.69,0.88,0.90}
\definecolor{purple1}{rgb}{0.61,0.19,1.00}
\definecolor{purple2}{rgb}{0.57,0.17,0.93}
\definecolor{purple3}{rgb}{0.49,0.15,0.80}
\definecolor{purple4}{rgb}{0.33,0.10,0.55}
\definecolor{purple}{rgb}{0.63,0.13,0.94}
\definecolor{red1}{rgb}{1.00,0.00,0.00}
\definecolor{red2}{rgb}{0.93,0.00,0.00}
\definecolor{red3}{rgb}{0.80,0.00,0.00}
\definecolor{red4}{rgb}{0.55,0.00,0.00}
\definecolor{red}{rgb}{1.00,0.00,0.00}
\definecolor{rosybrown}{rgb}{0.74,0.56,0.56}
\definecolor{royalblue}{rgb}{0.25,0.41,0.88}
\definecolor{saddlebrown}{rgb}{0.55,0.27,0.07}
\definecolor{salmon1}{rgb}{1.00,0.55,0.41}
\definecolor{salmon2}{rgb}{0.93,0.51,0.38}
\definecolor{salmon3}{rgb}{0.80,0.44,0.33}
\definecolor{salmon4}{rgb}{0.55,0.30,0.22}
\definecolor{salmon}{rgb}{0.98,0.50,0.45}
\definecolor{sandybrown}{rgb}{0.96,0.64,0.38}
\definecolor{seagreen}{rgb}{0.18,0.55,0.34}
\definecolor{seashell1}{rgb}{1.00,0.96,0.93}
\definecolor{seashell2}{rgb}{0.93,0.90,0.87}
\definecolor{seashell3}{rgb}{0.80,0.77,0.75}
\definecolor{seashell4}{rgb}{0.55,0.53,0.51}
\definecolor{seashell}{rgb}{1.00,0.96,0.93}
\definecolor{sienna1}{rgb}{1.00,0.51,0.28}
\definecolor{sienna2}{rgb}{0.93,0.47,0.26}
\definecolor{sienna3}{rgb}{0.80,0.41,0.22}
\definecolor{sienna4}{rgb}{0.55,0.28,0.15}
\definecolor{sienna}{rgb}{0.63,0.32,0.18}
\definecolor{skyblue}{rgb}{0.53,0.81,0.92}
\definecolor{slateblue}{rgb}{0.42,0.35,0.80}
\definecolor{slategray}{rgb}{0.44,0.50,0.56}
\definecolor{slategrey}{rgb}{0.44,0.50,0.56}
\definecolor{snow1}{rgb}{1.00,0.98,0.98}
\definecolor{snow2}{rgb}{0.93,0.91,0.91}
\definecolor{snow3}{rgb}{0.80,0.79,0.79}
\definecolor{snow4}{rgb}{0.55,0.54,0.54}
\definecolor{snow}{rgb}{1.00,0.98,0.98}
\definecolor{springgreen}{rgb}{0.00,1.00,0.50}
\definecolor{steelblue}{rgb}{0.27,0.51,0.71}
\definecolor{tan1}{rgb}{1.00,0.65,0.31}
\definecolor{tan2}{rgb}{0.93,0.60,0.29}
\definecolor{tan3}{rgb}{0.80,0.52,0.25}
\definecolor{tan4}{rgb}{0.55,0.35,0.17}
\definecolor{tan}{rgb}{0.82,0.71,0.55}
\definecolor{thistle1}{rgb}{1.00,0.88,1.00}
\definecolor{thistle2}{rgb}{0.93,0.82,0.93}
\definecolor{thistle3}{rgb}{0.80,0.71,0.80}
\definecolor{thistle4}{rgb}{0.55,0.48,0.55}
\definecolor{thistle}{rgb}{0.85,0.75,0.85}
\definecolor{tomato1}{rgb}{1.00,0.39,0.28}
\definecolor{tomato2}{rgb}{0.93,0.36,0.26}
\definecolor{tomato3}{rgb}{0.80,0.31,0.22}
\definecolor{tomato4}{rgb}{0.55,0.21,0.15}
\definecolor{tomato}{rgb}{1.00,0.39,0.28}
\definecolor{turquoise1}{rgb}{0.00,0.96,1.00}
\definecolor{turquoise2}{rgb}{0.00,0.90,0.93}
\definecolor{turquoise3}{rgb}{0.00,0.77,0.80}
\definecolor{turquoise4}{rgb}{0.00,0.53,0.55}
\definecolor{turquoise}{rgb}{0.25,0.88,0.82}
\definecolor{violetred}{rgb}{0.82,0.13,0.56}
\definecolor{violet}{rgb}{0.93,0.51,0.93}
\definecolor{wheat1}{rgb}{1.00,0.91,0.73}
\definecolor{wheat2}{rgb}{0.93,0.85,0.68}
\definecolor{wheat3}{rgb}{0.80,0.73,0.59}
\definecolor{wheat4}{rgb}{0.55,0.49,0.40}
\definecolor{wheat}{rgb}{0.96,0.87,0.70}
\definecolor{whitesmoke}{rgb}{0.96,0.96,0.96}
\definecolor{white}{rgb}{1.00,1.00,1.00}
\definecolor{yellow1}{rgb}{1.00,1.00,0.00}
\definecolor{yellow2}{rgb}{0.93,0.93,0.00}
\definecolor{yellow3}{rgb}{0.80,0.80,0.00}
\definecolor{yellow4}{rgb}{0.55,0.55,0.00}
\definecolor{yellowgreen}{rgb}{0.60,0.80,0.20}
\definecolor{yellow}{rgb}{1.00,1.00,0.00}

\def\Slash#1{{#1\!\!\!\slash}}

\def\nslash{n\!\!\!\slash}
\def\bnslash{\bar n\!\!\!\slash}

\def\OMIT#1{}

\newcommand{\req}[1]{Eq.\:\eqref{#1}}

\newcommand{\nn}{\nonumber} 

\newcommand{\bn}{{\bar n}}
\newcommand{\beq}{\begin{equation}}
\newcommand{\eeq}{\end{equation}}
\newcommand{\bea}{\begin{eqnarray}}
\newcommand{\eea}{\end{eqnarray}}

\newcommand{\mcdot}{\!\cdot\!}

\newcommand{\LQCD}{\mbox{$\Lambda_{\rm  had}$ }}

\newcommand{\SCETa}{\mbox{${\rm SCET}_{\rm I}$ }}
\newcommand{\SCETb}{\mbox{${\rm SCET}_{\rm II}$ }}

\begin{document}

%%%%%%%%%%%%%%%%%%%%%%%%%%%%%%%%%%%%%%%%%%
%Define Title, Author, Address, Preprint#

\title{Rapidity Divergences and Deep Inelastic Scattering in the Endpoint Region}

\author{Sean Fleming\footnote{Electronic address: fleming@physics.arizona.edu}}
  \affiliation{University of Arizona, Tucson, AZ 85721, USA}
  
\author{Ou Z. Labun\footnote{Electronic address: ouzhang@email.arizona.edu}}
 \affiliation{University of Arizona, Tucson, AZ 85721, USA}  

\date{\today\\ \vspace{1cm} }

%%%%%%%%%%%%%%%%%%%%%%%%%%%%%%%%%%%%%%%%%%
%Create the title page

\begin{abstract}
The deep inelastic scattering cross section in the endpoint region, $x \sim 1$, has been subjected to extensive analysis.
We revisit this process using soft collinear effective theory, and show that in the endpoint individual factors in the factorized hadronic tensor have rapidity divergences. We regulate these divergences using a recently introduced rapidity regulator, and find that each operator matrix element requires a different scale to minimize large rapidity logarithms. Unfortunately, the running in rapidity is non-perturbative and must be absorbed into the definition of the parton distribution function. 
\end{abstract}

\maketitle

%%%%%%%%%%%%%%%%%%%%%%%%%%%%%%%%%%%%%%%%%%
\newpage
%%%%%%%%%%%%%%%%%%%%%%%%%%%%%%%%%%%%%%%%%%
%Main body of the paper

\section{Introduction}

Deep inelastic scattering (DIS)  has been crucial in developing our understanding of QCD since the first high energy experiments at the Stanford linear accelerator in 1967~\footnote{For a nice review of the history of DIS see the Nobel lecture by Henry W. Kendall~\cite{kendallnobel}}. These early experiments gave rise to Feynman's parton model, and subsequent DIS experiments have allowed us to further refine our  understanding of the structure of nucleons. In this paper we explore DIS in a corner of phase space, where the light-cone momentum fraction, $x$, of the struck quark nears its maximal value, $x \sim1$. Ours is not the first analysis that has scrutinized this endpoint regime. 
Factorization and resummation of the DIS cross section for $x \sim 1$ was first investigated in Refs.~\cite{Sterman:1986aj,Korchemsky:1988pn,Catani:1989ne,Korchemsky:1992xv} using QCD factorization methods. Later, with the development of  soft collinear effective theory (SCET)~\cite{Bauer:2000ew,Bauer:2000yr,Bauer:2001yt}, DIS in the endpoint region was revisited in the context of effective field theory~\cite{Manohar:2003vb,Pecjak:2005uh,Chay:2005rz,Manohar:2005az,Idilbi:2006dg,Chen:2006vd,Becher:2006mr,Idilbi:2007ff}. 

In this work we use SCET to study the $x\sim 1$ region of DIS, and focus on the definition of each term in the factorized form of the hadronic tensor. We repeat the derivation of the factorization of the DIS hadronic tensor into a hard coefficient, a jet function, a collinear factor, and a soft function. Each of these pieces is well defined in SCET. The hard coefficient comes from the matching of SCET onto QCD, while the jet function, collinear factor, and soft function are matrix elements of SCET operators. The jet function consists of all radiation that is collinear to the final state, while the collinear factor consists of all radiation collinear to the initial state. The soft function includes soft radiation from both the initial and final state. Though the properties of the hard coefficient and jet function are well known, the collinear factor and soft function have not been explored as throughly, and it is on these latter two objects that we focus our attention. 

The collinear factor and soft function can be combined into a single non-perturbative parton distribution function (PDF) as was done in Ref.~\cite{,Pecjak:2005uh}. While, as we will argue, this is a sensible procedure there is something to be learned from considering the renormalization properties of the soft and collinear pieces separately: namely that combining these objects results in a single logarithm of widely mismatched rapidity scales. We carry out a one-loop calculation of the collinear and soft operator matrix elements using the rapidity regulator introduced in Refs.~\cite{Chiu:2011qc,Chiu:2012ir}. Our calculation explicitly shows that the collinear factor and the soft function each have a rapidity divergence and an associated logarithm of the rapidity scale $\nu$ that is minimized at $\nu \sim Q$ (where $Q$ is a large energy scale) for the collinear factor and at $\nu \sim Q(1-x) \ll Q$ for the soft function. When the soft function and collinear factor are combined into the PDF the rapidity divergences cancel, however, a single large logarithm of the ratio of collinear and soft rapidity scales is left over. This large logarithm shows up both in the finite part of the one-loop expression for the PDF, and in the PDF anomalous dimension. This is the first time that the presence of a single large logarithm in the endpoint region of DIS has been identified and explained; it is one of the main results of this paper.

Our calculations have a number of interesting aspects. First, SCET label momentum conservation and the collinear zero-bin subtraction~\cite{Manohar:2006nz} forces real emission from the initial state to be soft~\cite{Idilbi:2007ff}, which is a characteristic that distinguishes end-point DIS from DIS at moderate $x$. As was first noted in Ref.~\cite{Becher:2006mr}, this implies that the collinear factor has only virtual contributions. Second, since the soft and collinear functions are both described in $\textrm{SCET}_{\rm{ II}},$ there exists a soft zero-bin in which any overlap of the soft degrees of freedom with collinear ones also has to be subtracted from the soft contributions. In this work, we present the first computation of these zero-bin subtractions using the rapidity regulator of Ref.~\cite{Chiu:2011qc,Chiu:2012ir}. Third, we find that the choice of scale which minimizes large rapidity logarithms is different in the collinear factor and soft function so that a resummation of rapidity logarithms is needed. This running in rapidity is, unfortunately, non-perturbative, and must therefore be absorbed into the non-perturbative soft function. This implies that a model of the PDF in the endpoint might need to include logarithmically enhanced parameters. Finally, we show that the soft function, which naively is expressed in terms of soft Wilson lines extending from the initial state into the final state, can be expressed only in terms of Wilson lines in the initial state. This guarantees the universality of the PDF in the sense that it only depends on the initial hadronic state.

\section{Factorization}

In this section, we use SCET to repeat the derivation of the factorization of the DIS hadronic tensor. We work in the Breit frame where the incoming proton moves along the $-\hat{z}$ direction with energy much larger than the proton mass $m_p$, so that the proton momentum is 
\beq
p^\mu = \frac{\sqrt{s}}{2}n^\mu + \frac{m^2_p}{2 \sqrt{s}}\bn^\mu\,,
\eeq
where $n^\mu= (1,0,0,-1)$, $s= (p+k)^2$,  is the center-of-mass energy squared, and $m_p$ is the proton mass.
Particles collinear to the proton have momentum
\beq
p^\mu_n =  \frac{1}{2} \bn\cdot p_n \, n^\mu + \frac{1}{2} n\cdot p_n \,  \bn^\mu+p^\mu_{n,\perp}\,,
\eeq
where components differ parametrically in their sizes:
$\bn\cdot p_n \sim \sqrt{s}$, $p^\mu_{n,\perp} \sim \LQCD/\sqrt{s}$,  and $n\cdot p_n \sim (\LQCD/\sqrt{s})^2$, with $\LQCD\sim m_p$  a typical hadronic scale. 
The incoming proton is struck by a virtual gluon of momentum $q^\mu$  with large invariant mass squared: $-q^2 \equiv Q^2$. The final state momentum is restricted by momentum conservation to be $p_X = p+q$ with invariant mass squared
\beq
M^2_X = (p+q)^2 =\frac{Q^2}{x}(1-x)+m_p^2\approx \frac{Q^2}{x}(1-x) \,, \qquad\qquad x = \frac{Q^2}{2p_p\cdot q} \,.
\eeq
In the endpoint region we are considering the invariant mass $M_{x}\approx Q\sqrt{1-x}$ is  small compared to $Q$, but is much larger than the typical hadronic scale $\Lambda_{\rm{had}}$. Note, we do not fix the scale $Q(1-x)$ relative to $\Lambda_{\rm{had}}$.
Thus the total final state momentum in the endpoint region is collinear, and any final state collinear particle will have momentum 
\beq
p^\mu_\bn =  \frac{1}{2} n \cdot p_\bn \, \bn^\mu + \frac{1}{2} \bn \cdot p_\bn \, n^\mu + p^\mu_{\bn, \perp}\,,
\eeq
where $\bn ^{\mu}=(1,0,0,1)$, $n \cdot p_\bn \sim Q$ , $p^\mu_{\bn,\perp} \sim Q\lambda$ and $ \bn \cdot p_\bn \sim Q \lambda^2$ with $\lambda \sim \sqrt{1-x}$.
Finally, we have 
\beq
\label{momenta}
q^\mu =  \frac{Q}{2} \, (\bn^\mu -n^\mu)\,,
\eeq
and
\beq
\sqrt{s} = Q+Q\frac{1-x}{x}+...\approx Q
\eeq
which means that up to correction of order $\lambda^2$ the large lightcone momentum component of the proton is 
\beq
\label{protlabel}
\bn \cdot p \approx  Q \,.
\eeq

In our analysis we follow a two-step procedure: in the first step we match from QCD onto \SCETa where the offshellness of collinear 
momentum scales as $p^2_c \sim Q^2 \lambda^2$, in the next step we integrate out the final state collinear fields and match onto \SCETb where the offshellness of 
collinear fields scale as $p^2_c \sim \Lambda^{2}_{\rm{had}}$. The first step is straightforward and has been covered in detail in Ref.~\cite{Manohar:2003vb}. In this work we are only concerned with the second one. 
Following Ref.~\cite{Bauer:2002nz},  the DIS cross section is 
\beq
\sigma = \frac{d^3 \mathbf{k}'}{2 |{\mathbf{k}'|} (2 \pi)^3}\, \frac{\pi e^4}{s Q^4} L_{\mu \nu}(k,k') W^{\mu \nu}(p,q)\,,
\end{equation}
where $k$ and $k'$ are the incoming and outgoing lepton momenta, $q = k-k'$, and
\begin{equation}
L_{\mu \nu} = 2( k_\mu k'_\nu +  k_\nu k'_\mu- k\cdot k' g_{\mu \nu}) \,.
\eeq
The DIS hadronic tensor is
\beq
\label{htensor}
 W^{\mu \nu}(p,q)   =  \frac{1}{2} \sum_{\sigma}  \int d^4x \, e^{iq\cdot x}\, \langle h( p,\sigma)|  J^{\mu}(x) J^\nu(0)  |h(p,\sigma) \rangle  \,,
\eeq
with 
\beq
 J^\mu(x) =  \bar{\psi}(x) \gamma^\mu \psi(x) \,,
\eeq
and external proton state $h(p,\sigma)$ with momentum $p$ and spin $\sigma$.
The QCD current in Eq.~(\ref{htensor}) matches onto an SCET current of the form
\beq
J^\mu_{\rm eff} (x)= \bar{\chi}_{\bn, \omega_2} \gamma^\mu_\perp \chi_{n, \omega_1}(x)  + H.c. \,,
\eeq
where $H.c.$ stands for the hermitian conjugate. Matching gives
\beq
J^\mu (x) \to  \sum_{\omega_1, \omega_2} C(\omega_1, \omega_2;\mu_q, \mu)\bigg(e^{-\frac{i}{2} \omega_1 n \cdot x} e^{\frac{i}{2} \omega_2 \bn\cdot x} \bar{\chi}_{\bn, \omega_2} \gamma^\mu_\perp \chi_{n, \omega_1}(x)  + h.c. \bigg)\,,
\eeq
where 
\beq
\gamma^\mu_\perp \equiv \gamma^\mu - \frac{1}{2}\nslash \bn^\mu-\frac{1}{2}\bnslash n^\mu\,,
\eeq
and the coefficient $C(\omega_1, \omega_2;\mu_q, \mu)$ depends on a factorzation scale $\mu_q$ at which the matching onto QCD is carried out, and a running scale $\mu$. 
From Eq.~(\ref{htensor}) we determine the hadronic tensor in \SCETa:
\bea
\label{htscet}
W^{\mu \nu}_{\rm eff} &=& \sum_{\omega_1, \omega_2, \omega'_1, \omega'_2} 
C^*(\omega_1,\omega_2;\mu_q, \mu) C(\omega'_1, \omega'_2;\mu_q, \mu)  \int \frac{d^4x}{4 \pi} \,  
e^{-\frac{i}{2}(Q -\omega_1)n \cdot x} e^{\frac{i}{2}(Q -\omega_2)\bn \cdot x}e^{-\frac{i}{2}Q\frac{1-x}{x}n\cdot x} \nn \\
& &\qquad\times \frac{1}{2} \sum_\sigma \sum_{\bn\cdot\tilde p} \delta_{\bn\cdot\tilde p, Q}\langle{h_{n}(p,\sigma)}|
\bar{\textrm{T}}\bigg[ \bar{\chi}_{n, \omega_1} \gamma^\mu_\perp \chi_{\bn, \omega_2}(x)\bigg]
\textrm{T}\bigg[ \bar{\chi}_{\bn, \omega'_2} \gamma^\nu_\perp \chi_{n, \omega'_1}(0)\bigg] 
|{h_{n}(p,\sigma)}\rangle \nn \\
&=&  \sum_{\omega_1, \omega_2, \omega'_1, \omega'_2} \delta_{Q ,\omega_1} \delta_{Q,\omega_2}
C^*(\omega_1,\omega_2;\mu_q, \mu) C(\omega'_1, \omega'_2;\mu_q, \mu)  \int \frac{d^4 x}{4 \pi}e^{-\frac{i}{2}Q\frac{1-x}{x} n\cdot x}  \nn \\
& & \qquad\times \frac{1}{2} \sum_\sigma  \sum_{\bn\cdot\tilde p} \delta_{\bn\cdot\tilde p,  Q}\langle{h_{n}(p,\sigma)}|
 \bar{\textrm{T}}\bigg[\bar{\chi}_{n, \omega_1} \gamma^\mu_\perp \chi_{\bn, \omega_2}(x)\bigg]
 \textrm{T}\bigg[  \bar{\chi}_{\bn, \omega'_2} \gamma^\nu_\perp \chi_{n, \omega'_1}(0)\bigg] 
 |{h_{n}(p,\sigma)}\rangle \,,\nn\\
\eea
where $\textrm{T}$ denotes time ordering, $\bar{\textrm{T}}$ anti-time ordering, and $h_{n}(p,\sigma)$ denotes the SCET proton state.
Here we have inserted a sum over proton label momentum and an explicit Kronecker delta that ensures the proton label momentum is equal to $ Q$ as required by momentum conservation, Eq.~(\ref{protlabel}).
Usoft gluons in \SCETa can be decoupled from collinear modes via the BPS phase redefinition~\cite{Bauer:2001yt}, and the hadronic tensor above can be factored into matrix elements of operators in each of the two collinear sectors and the usoft sector:
\bea
\label{factoredht}
W^{\mu \nu}_{\rm eff} &=& \frac{- g_\perp^{\mu\nu}}{2} N_c 
 \sum_{\omega'_1, \omega'_2} C^*(Q ,Q;\mu_q, \mu ) C(\omega'_1,\omega'_2;\mu_q, \mu)\int \frac{d^4 x}{4 \pi}e^{-\frac{i}{2}Q\frac{1-x}{x}n\cdot x}
 \\
&\times&  \frac{1}{2} \sum_\sigma \sum_{\bn\cdot\tilde p}  \delta_{\bn\cdot\tilde p, Q} \langle{h_{n}(p,\sigma)}|  \bar{\chi}_{n, Q}(x)\frac{\bnslash}{2} \chi_{n, \omega'_1}(0)|{h_{n}(p,\sigma)}\rangle \nn\\
&\times& \langle{0} |\frac{\nslash}{2}\chi_{\bn, Q}(x) \bar{\chi}_{\bn, \omega'_2}(0) |0\rangle \nn \\
&\times& \frac{1}{N_c} \langle{0} |\textrm{Tr}\bigg(
\bar{\textrm{T}}\bigg[ Y^\dagger_{n}(x) \tilde{Y}_{\bn}(x)\bigg]
\textrm{T}\bigg[ \tilde{Y}^\dagger_{\bn}(0) Y_{n}(0) \bigg]
\bigg)|0\rangle \,.\nn
\eea
The Wilson lines $Y_n$ and $\tilde Y_{\bn}$ associated with soft radiation from the initial and final state respectively are defined as
\bea
Y_n(x) &=& \textrm{P exp}\bigg( i g \int^x_{-\infty} ds\, n\!\cdot\! A_{us}(sn)\bigg)   \\
\tilde Y_{\bn}(x) &=&\textrm{P exp}\bigg( i g \int^{\infty}_x ds\,\bn\!\cdot\! A_{us}(s\bn)\bigg)\,.   \nn
\eea
We define a jet function 
\bea
\label{jetfun}
&&  \langle{0} |\frac{\bnslash}{2}\chi_{\bn, \omega_2}(x) \bar{\chi}_{\bn, \omega'_2}(0)|0 \rangle
\equiv   Q  \delta(\bn\cdot x) \delta^{(2)}(x_{\perp}) \int dr \, e^{-\frac{i}{2}r n \cdot x} J_{\bn}(r;\mu) \,,
\eea
and a soft function
\bea
\label{softfun}
&&\frac{1}{N_c} \langle{0} |\textrm{Tr}\bigg(
\bar{\textrm{T}}\bigg[ Y^\dagger_{n}(n\cdot x) \tilde{Y}_{\bn}(n\cdot x) \bigg]
\textrm{T}\bigg[\tilde{Y}^\dagger_{\bn}(0) Y_{n}(0) \bigg]
\bigg)|0\rangle
 \equiv  \int d\ell \, e^{-\frac{i}{2}\ell n\cdot x} S(\ell;\mu) \,.
\eea
Then, we use label momentum conservation to simplify the collinear matrix element in the $n$ sector:
\bea
\label{colsimp}
& &  \langle{h_{n}(p,\sigma)} |\bar{\chi}_{n,  Q}(x)\frac{\bnslash}{2} \chi_{n, \omega'_1}(0)|{h_{n}(p,\sigma)}\rangle \nn \\
&& \hspace{5 ex}=\delta_{ Q,\omega'_1} \,\langle{h_{n}(p,\sigma)}| \bar{\chi}_{n}(x)\frac{\bnslash}{2} \delta_{\bar{\cal P}, 2  \bn\cdot\tilde p}  \chi_{n}(0) |{h_{n}(p,\sigma)}\rangle \,,
\eea
where $\bar{\cal P}= \bn\mcdot({\cal P}+{\cal P}^\dagger)$ projects out label momentum~\cite{Bauer:2001ct}, and $\bn\cdot\tilde p$ is the large component of the proton momentum.
Using the definitions in Eqs.~(\ref{jetfun},\ref{softfun}) and the relation in Eq.~(\ref{colsimp}), the hadronic tensor in Eq.~(\ref{factoredht}) becomes
\bea
\label{hadten1}
W^{\mu \nu}_{\rm eff} &=&   -g_\perp^{\mu\nu}
H (Q;\mu_q, \mu ) \int dr d \ell \, J_{\bn}(r;\mu) S(\ell;\mu) \int \frac{d\, n\mcdot x}{4 \pi} \, e^{-\frac{i}{2}(r+\ell+Q\frac{1-x}{x})n\cdot x}\nn \\
&&\qquad \times \frac{1}{2}\sum_\sigma  \sum_{\bn\cdot\tilde p} 
\delta_{\bn\cdot\tilde p,  Q} \, \langle{h_{n}(p,\sigma)} | \bar{\chi}_{n}(n\mcdot x)\frac{\bnslash}{2} \delta_{\bar{\cal P}, 2  Q} \chi_{n}(0) |{h_{n}(p,\sigma)}\rangle \,, 
\eea
where 
\beq
H( Q; \mu_{q},\mu ) =  Q |C( Q , Q ;\mu_q, \mu) |^2 \,.
\eeq
Finally, we introduce an $n$-collinear function
\bea
\label{n1collfun}
{\cal C}_{n}( k;\mu)  
&&= \int \frac{d\, n\mcdot x }{4 \pi}\, e^{\frac{i}{2}kn\cdot x} \frac{1}{2} \sum_\sigma  \sum_{\bn\cdot\tilde p} 
\delta_{\bn\cdot\tilde p,  Q}\, \langle{h_{n}(p,\sigma)}| \bar{\chi}_{n}(n\mcdot x)\frac{\bnslash}{2} \delta_{\bar{\cal P}, 2  Q} \chi_{n}(0) |{h_{n}(p,\sigma)}\rangle \nn \\
&&=\frac{1}{2} \sum_\sigma   \sum_{\bn\cdot\tilde p} 
\delta_{\bn\cdot\tilde p,  Q}\, \langle{h_{n}(p,\sigma)}| \bar{\chi}_{n}(0)\frac{\bnslash}{2} \delta_{\bar{\cal P}, 2  Q} \delta(i\bn\cdot\partial -k) \chi_{n}(0) |{h_{n}(p,\sigma)}\rangle\,.
\eea
Using this definition in Eq.~(\ref{hadten1}) we arrive at our final expression for the factored form of the DIS hadronic tensor in \SCETa:
\bea
\label{hadtenfin}
W^{\mu \nu}_{eff} &=& -g^{\mu\nu}_\perp H( Q ;\mu_q, \mu)   
 \int dr d \ell \, J_{\bn}(r;\mu) S(\ell;\mu) {\cal C}_{n}( Q+r+\ell;\mu) \,.\nn\\
 &&
\eea
The $\mu$ dependence of the hard coefficient $H$ is  such that it exactly cancels the $\mu$ dependence of the product of the collinear and soft functions.

It is now straightforward to match Eq.~(\ref{hadtenfin}) onto \SCETb. The jet function $J_{\bn}(r;\mu)$ characterizes the final state with typical offshellness $M_x^2 \sim Q^2(1-x)$, and can be integrated out at the scale $\mu_c \sim Q\sqrt{1-x}$. The usoft gluons of \SCETa become soft gluons in $\textrm{SCET}_{\rm{II}},$ so $S(\ell;\mu)$ remains unchanged. The off-shellness of the $n$-collinear degrees of freedom changes from $p^2_c \sim Q^2 (1-x)$ in \SCETa to $p^2_c \sim \Lambda^{2}_{\rm{had}} $ in \SCETb and ${\cal C}_{n}$ also remains unchanged. As was pointed out in Refs.~\cite{Chiu:2011qc,Chiu:2012ir} the factorization of soft and collinear modes in \SCETb requires an additional regulator which separates rapidity regions, so $S$ and ${\cal C}_{n}$  will depend on a rapidity scale that cancels between the two. Since there can be no collinear radiation into the final state in the $x \sim1$ region the collinear function can be expressed as
\beq
 {\cal C}_{n}( k;\mu,\nu)={\cal Z}_{n}(Q;\mu,\nu) \, \delta(k)\,,
\eeq 
where $\nu$ plays the role of a dimensionful rapidity scale separating soft and collinear rapidity regions.
Thus, in \SCETb the hadronic tensor is
\beq
\label{hadtenSCET2}
W^{\mu \nu}_{eff} = -g_\perp^{\mu\nu}H( Q ;\mu_q, \mu_c)   
 \int d \ell \, J_{\bn}(\ell;\mu_c,\mu) \phi^{ns}_q(Q\frac{1-x}{x}+\ell;\mu) ,
\eeq
with
\beq
\label{phi}
\phi^{ns}_q(\ell;\mu) =  {\cal Z}_{n}(Q;\mu,\nu)S(\ell; \mu, \nu)\,,
\eeq
defining the non-perturbative parton distribution function. The scale that minimizes rapidity logarithms in ${\cal Z}_{n}$ is different from the scale that minimizes rapidity logarithms in $S$. However, the $\nu$ dependence cancels on the right-hand side so the PDF is $\nu$ independent. Our expression agrees with the expression in Ref.~\cite{Becher:2006mr} up to the appearance of the rapidity regulator which was not considered in that work. One may worry that identifying $\phi^{ns}_q$ with the PDF is problematic because $\phi^{ns}_q$ depends on final state soft radiation. However, as we show in 
Sect.~\ref{pdf}, $\phi^{ns}_q$ can be expressed only in terms of initial state Wilson lines, which ensures the universality of the PDF.

\section{The collinear function}

In this section we study the collinear function ${\cal C}_{n}( k;\mu,\nu)$. As was first done in Ref.~\cite{Idilbi:2007ff}, we argue that label momentum conservation and the zero bin subtraction allows no real radiation of $n$-collinear particles into the final state so this function involves only virtual corrections. We explicitly show how this works at one loop in perturbation theory. In addition, up to the same order we show the need for a rapidity regulator and determine the value of the rapidity scale $\nu$ which minimizes rapidity logarithms.

The label momentum conserving Kronecker delta $\delta_{\bn\cdot\tilde p,  Q}$ in Eq.~(\ref{n1collfun}) forces the external proton label momentum  to be equal to $ Q$, while the Kronecker delta $\delta_{\bar{\cal P}, 2  Q}$ requires that each $\chi_{n}$ field has total label moment $ Q$ as well. Thus, any momentum that flows from the $\bar{\chi}_{n}$ field on the left side to the $\chi_{n}$ field on the right must have zero label momentum. Any field that causes momentum to flow in this way corresponds to real radiation (as it must cross the cut). Since SCET is formulated with an explicit zero-bin subtraction, collinear fields with zero label momentum vanish, which means that there can be no real radiation of $n$-collinear particles. This is just a manifestation of momentum conservation: only soft radiation from the initial state into the final state is allowed otherwise we are no longer in the $x \sim 1$ region. 

Let us consider an explicit calculation of ${\cal C}_{n}( k;\mu,\nu)$ to order $\alpha_s$ using external parton states.  
The ${\cal O}(\alpha^{0}_s)$ Feynman diagram is shown in Fig.~\ref{colltreefig}
\begin{figure}[htbp]
  \centering
   \includegraphics[scale=0.6]{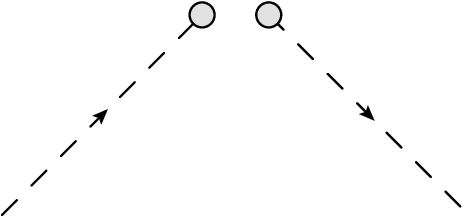} % requires the graphicx package
   \caption{The ${\cal O}(\alpha_s^{0})$ Feynman diagram for the $n$ collinear function. The dashed lines are collinear quarks, the grey circles are vertices where 
momentum is injected, and the gap indicates a cut.}
   \label{colltreefig}
\end{figure}
and gives the tree level result
\bea
{\cal C}_{n}( k)^{(0)}& = &  \sum_{\bn\cdot\tilde p}  \delta_{\bn\cdot\tilde p, Q}\delta_{\bn\cdot\tilde p, Q}\delta(\bn\cdot p_r -k) m_0\\
&=&  \delta(k) m_0 \,,\nn
\eea
where $\bn\cdot\tilde p$ is the ${\cal O}(1)$ quark label momentum, and $p_r$ is the ${\cal O}(\lambda^2)$ quark residual momentum, which can be set to zero for an on-shell quark. The two Kronecker deltas in the first line come directly from the definition of the operator in Eq.~(\ref{n1collfun}). Here
\beq
m_0 = \frac{1}{2}\sum_\sigma \bar \xi_{n}^\sigma \frac{\bnslash_1}{2}\xi_{n}^\sigma\,,
\eeq
where $\xi_{n}^\sigma$ are SCET quark spinors with spin $\sigma$.

Three of the five ${\cal O}(\alpha_s)$ Feynman diagram for ${\cal C}_{n}( k;\mu,\nu)$ are shown in Fig.~\ref{oneloopcoll}. The remaining two diagrams are obtained by the reflection of diagrams (a) and (b) about a vertical axis through the middle of the diagram.
\begin{figure}[htbp]
   \centering
   \includegraphics[scale=0.95]{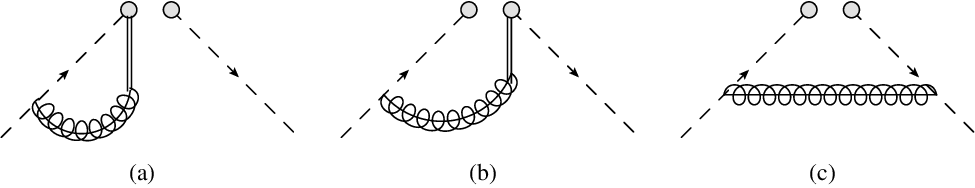} % requires the graphicx package
   \caption{The one-loop Feynman diagrams for the $n$ collinear function: (a) is the virtual contribution, while (b) and (c) are real contributions. Two diagrams which are the mirror image of (a) and (b) are not shown. The double line represents a Wilson line which is the source of a single gluon.  }
   \label{oneloopcoll}
\end{figure}
The  amplitude corresponding to diagram (a) is
\beq
\label{ma}
im_{\textrm{(a)}} = i m_0 \, (2 g_s^2 C_F) \sum_{\bn\cdot\tilde p} \delta_{\bn\cdot \tilde p, Q}\delta(\bn\cdot p_r-k) \, \mu^{2 \epsilon} \sum_{\bn\cdot \tilde q \neq 0} \int \frac{d^Dq_r}{(2\pi)^D} \frac{1}{\bn\cdot q}\frac{\bn\cdot(p-k)}{(p-q)^2+i \epsilon}
\frac{1}{q^2+i \epsilon} 
\eeq
where we work in $D= 4 -2 \epsilon$ dimensions, and the external quark states have momentum $p^\mu$. This diagram gives a virtual correction since the gluon does not cross the cut. The sum over the gluon label momentum is restricted to those values where $\bn\cdot\tilde q \neq 0$ to prevent double counting of degrees of freedom~\cite{Manohar:2006nz}. 
The amplitude obtained from diagram (b) is
\bea
i m _{\textrm{(b)}}&=&-i m_0 \, (2 g_s^2 C_F) \sum_{\bn\cdot\tilde p} \delta_{\bn\cdot \tilde p, Q} \, \mu^{2 \epsilon}\\
&& \hspace{5ex} \times \sum_{\bn\cdot \tilde q \neq 0}  \delta_{\bn\cdot \tilde q,0}\int \frac{d^Dq_r}{(2\pi)^{D-1}} \frac{1}{\bn\cdot q}\frac{\bn\cdot(p-k)}{(p-q)^2+i \epsilon}
\delta(q^2)\delta(\bn\cdot p_r-k) \,, \nn
\eea
and corresponds to real radiation as the gluon crosses the cut. Note, because of label momentum conservation the real gluon must have zero label momentum, as enforced by the $\delta_{\bn\cdot \tilde q,0}$. Since all collinear fields are defined such that $\bn\cdot \tilde q \neq 0$, $i m _{\textrm{(b)}}= 0$ and there is no real collinear radiation in the amplitude. Similarly, $i m _{\textrm{(c)}}= 0$. 

Including the contribution from the reflected diagrams which are not shown in  Fig.~\ref{colltreefig} the total collinear contribution will be twice that in Eq.~(\ref{ma}).
\beq
\label{colloneloop}
{\cal C}_{n}( k)^{(1)}  =  m_0\sum_{\bn\cdot\tilde p}  \delta_{\bn\cdot \tilde p, Q} \delta(k) \, (4 g_s^2 C_F) \, \mu^{2 \epsilon}  \int_{[\Slash{0}]} \frac{d^Dq}{(2\pi)^D} \frac{1}{\bn\cdot q}\frac{\bn\cdot(p-k)}{(p-q)^2+i \epsilon}\frac{1}{q^2+i \epsilon} \,,
\eeq
where the $[\Slash{0}]$ subscript indicates that the integral requires a zero-bin subtraction.
As was thoroughly discussed in Ref.~\cite{Chiu:2012ir} the integral in Eq.~(\ref{colloneloop}) contains a rapidity divergence that must be regulated properly. Here we adopt the approach in Ref.~\cite{Chiu:2012ir}, and introduce a gluon mass to regulate infrared (IR) divergences. Then, in agreement with Ref.~\cite{Chiu:2012ir} we find 
\bea
\label{finalcoll}
{\cal C}_{n}( k)^{(1)} &=&  m_0 \sum_{\bn\cdot\tilde p} \delta_{\bn\cdot \tilde p, Q} \delta(k) \, \frac{\alpha_s C_F}{\pi}w^2\bigg\{ \frac{e^{\epsilon \gamma_E} \Gamma(\epsilon)}{\eta}\bigg(\frac{\mu^2}{m^2_g}\bigg)^\epsilon +\frac{1}{\epsilon}\bigg[ 1 + \ln\frac{\nu}{\bn\cdot p} \bigg]  \\
&&\hspace{30ex}+ \ln\frac{\mu^2}{m^2_g}\ln\frac{\nu}{\bn\cdot p}+\ln\frac{\mu^2}{m^2_g}+1 -\frac{\pi^2}{6}\bigg\} \,. \nn 
\eea
Here $\eta$ is the rapidity regulator and $\nu$ the running rapidity scale. Clearly the logarithms in the expressions are minimized for a choice $\nu \sim \bn\cdot p\approx Q$ and $\mu \sim \Lambda_{\rm{had}}$. The divergences in $\eta$ and $\epsilon$ must be absorbed into appropriate counter-terms, as we discuss in Sect.~\ref{rnr}.

\section{The soft function}
Next we turn our attention to the soft function defined in Eq.~(\ref{softfun}). Our aim is to calculate the soft function to one loop so that we can isolate the poles in $\eta$, and determine the scale which minimizes rapidity logarithms.  At tree level we have the trivial result
\beq
S(\ell)^{(0)} = \delta(\ell) \,.
\eeq 
The one loop result is given by the sum of the diagrams in Fig.~\ref{softoneloop} and their reflections about a vertical axis through the middle of the diagram. The gap between the vertices indicates a cut in the diagram, so diagram (a) corresponds to a virtual contribution, while diagram (b) corresponds to a real contribution.
\begin{figure}[htbp]
   \centering
   \includegraphics[scale=0.75]{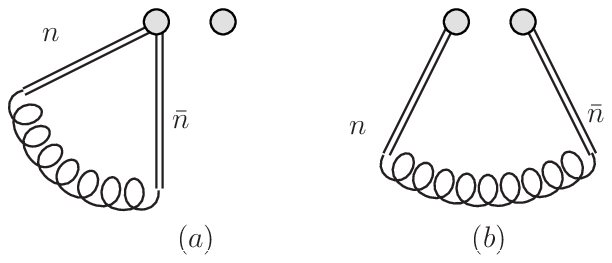} 
   \caption{Feynman diagrams for the one-loop evaluation of the soft function: (a) is the virtual contribution and (b) is the real contribution. There are two additional diagrams which are obtained by reflecting about a vertical axis through the middle of the diagram. The double lines indicate Wilson lines which produce the gluons, and $n$ and $\bar n$ label the direction of the Wilson lines. The gap between vertices indicates a cut.}
   \label{softoneloop}
\end{figure}
Again, in agreement with Ref.~\cite{Chiu:2012ir} we obtain
\beq
m_\textrm{v} = \delta(\ell) \frac{2 \alpha_s C_F}{\pi}w^2\bigg[ -\frac{e^{\epsilon \gamma_E} \Gamma(\epsilon)}{\eta}\bigg(\frac{\mu}{m_g}\bigg)^{2\epsilon} 
+\frac{1}{2 \epsilon^2}+\frac{1}{\epsilon}\ln \frac{\mu}{\nu}
+\ln^2\frac{\mu}{m_g}- \ln\frac{\mu^2}{m_g^2}\ln\frac{\nu}{m_g} - \frac{\pi^2}{24}\bigg]
\eeq
for diagram (a) in Fig.~\ref{softoneloop}. The real contribution from diagram (b) is
\bea
\label{realint}
\tilde m_\textrm{r} &=& - 2 C_F g^2_s \mu^{2 \epsilon} w^2 \nu^\eta \int \frac{d^Dk}{(2\pi)^{D-1}}\delta(k^2-m^2_g) \delta(\ell-k^+)|2 k^3|^{-\eta} \frac{1}{k^+}\frac{1}{k^-}  \\
& &= -\frac{\alpha_s C_F}{2 \pi}\bigg(e^{ \gamma_E} \frac{\mu^2}{m^2_g}\bigg)^\epsilon w^2 \nu^\eta \frac{\theta(\ell)}{\ell^{1+\eta}} \Gamma(\epsilon)\,.\nn
\eea
As pointed out in Ref.~\cite{Manohar:2006nz} in \SCETb there are also zero bin subtractions for the soft modes: any overlap with the $n$-collinear or $\bn$-collinear region must be removed. The virtual contribution zero-bin subtractions are all scale free and therefore vanish. The real contribution zero-bin subtractions, however, are not zero because the measurement function introduces an external scale into the one-loop integrals. The overlap of the integral in Eq.~(\ref{realint}) with the $n$-collinear region is given by taking the limit $k^+\gg k^-$ with $k^+k^- \sim k^2_\perp$
\bea
s_{n} &=&  - 2 C_F g^2_s \mu^{2 \epsilon} w^2 \nu^\eta \int \frac{d^Dk}{(2\pi)^{D-1}}\delta(k^2-m^2_g) \delta(\ell-k^+)|k^+|^{-\eta} \frac{1}{k^+}\frac{1}{k^-}  \\
& &= -\frac{\alpha_s C_F}{2 \pi}\bigg(e^{ \gamma_E} \frac{\mu^2}{m^2_g}\bigg)^\epsilon w^2 \nu^\eta \frac{\theta(\ell)}{\ell^{1+\eta}} \Gamma(\epsilon)\,,\nn
\eea
which is the same as the result in Eq.~(\ref{realint}). The $\bn$-collinear subtraction is given by taking the limit $k^-\gg k^+$ with $k^+k^- \sim k^2_\perp$ in the first line of Eq.~(\ref{realint}):
\bea
s_{\bn} &=&  - 2 C_F g^2_s \mu^{2 \epsilon} w^2 \nu^\eta \int \frac{d^Dk}{(2\pi)^{D-1}}\delta(k^2-m^2_g) \delta(\ell-k^+)|k^-|^{-\eta} \frac{1}{k^+}\frac{1}{k^-}  \\
& &= -\frac{\alpha_s C_F}{2 \pi}\bigg(e^{ \gamma_E} \frac{\mu^2}{m^2_g}\bigg)^\epsilon w^2 \bigg(\frac{\nu}{m^2_g}\bigg)^\eta \frac{\theta(\ell)}{\ell^{1-\eta}} 
\frac{\Gamma(\eta+\epsilon)}{\Gamma(1+\eta)}\,,\nn
\eea
Thus the zero bin subtracted real contribution is
\bea
\label{softreal}
m_\textrm{r} &=& \tilde m_\textrm{r}-s_{n} -s_{\bn} = -s_{\bn} \\
&=& 2 \frac{\alpha_s}{\pi} w^2 \bigg\{\bigg[\frac{1}{2} \frac{e^{\epsilon \gamma_E} \Gamma(\epsilon)}{\eta}\bigg(\frac{\mu}{m_g}\bigg)^{2\epsilon}
-\frac{1}{2\epsilon^2} +\frac{1}{2 \epsilon} \ln\frac{\nu}{\mu^2}-\ln^2\frac{\mu}{m_g} + \ln\frac{\mu}{m_g}\ln\frac{\nu}{m^2_g}+\frac{\pi^2}{24}\bigg]\delta(\ell)\nn\\
&& \hspace{8ex}+\bigg[\frac{1}{2\epsilon}+\ln\frac{\mu}{m_g}\bigg] \frac{1}{\ell_+}\bigg\} \,, \nn
\eea
where the plus-function of the dimensionfull variable $\ell$ is given in terms of the definition of a dimensionless variable $x = \ell/\kappa$
\beq
\frac{1}{(\ell)_+} = \frac{1}{\kappa (x)_+} + \ln \kappa \, \delta(\kappa \,x) \,,
\eeq
with
\beq
\frac{1}{(x)_+} \equiv \lim_{\beta \to 0} \bigg[ \frac{\theta(x-\beta)}{x}+\ln \beta \, \delta(x)\bigg] \,.
\eeq

As was pointed out in Refs.~\cite{Idilbi:2007ff,Idilbi:2007yi} the net effect of the zero bin subtraction without a rapidity regulator is to divide by the square of the soft matrix element.\footnote{In the QCD factorization approach the need for subtracting soft/collinear overlap terms in the case of DIS endpoint divergences was noted in Ref. \cite{Hautmann:2007uw}, where an explicit subtraction procedure was carried out based on the method of Ref. \cite{Collins:1999dz}.} In perturbation theory this is equivalent to subtracting the soft contribution.  With the introduction of a rapidity regulator this equivalence no longer holds. The only non-zero zero-bin subtraction in the virtual pieces is from the overlap of $n$-collinear modes in Eq.~(\ref{ma}) with the soft region. This is equivalent to dividing by a single power of the  soft matrix element. There is, however, no real collinear contribution, and the zero-bin subtractions for the real part come from the overlap of the soft integral with the two collinear regions. These subtractions are not equivalent to dividing by the square of the soft function, and a more complex picture emerges once rapidity divergences are isolated.
Adding the virtual and real contributions and multiplying by two to account for the mirror image diagrams gives the one loop expression for the soft function
\bea
S(\ell)^{(1)} &=& \frac{\alpha_s C_F}{\pi}w^2\bigg\{ -\frac{e^{\epsilon \gamma_E} \Gamma(\epsilon)}{\eta}\bigg(\frac{\mu}{m_g}\bigg)^{2\epsilon} \delta(\ell)
 +\bigg( \frac{1}{\epsilon}+ \ln\frac{\mu^2}{m^2_g}\bigg)\bigg[\frac{1}{(\ell)_+}- \ln{\nu}\, \delta(\ell)\bigg]\bigg\}\,.
\eea
When this expression is written in terms of the dimensionless variable $z = \ell/\kappa$ we find 
\bea
\label{finalsoft}
S(\ell)^{(1)} &=& \frac{\alpha_s C_F}{\pi}w^2\bigg\{ -\frac{e^{\epsilon \gamma_E} \Gamma(\epsilon)}{\eta}\bigg(\frac{\mu}{m_g}\bigg)^{2\epsilon} \delta(z)
 +\bigg( \frac{1}{\epsilon}+ \ln\frac{\mu^2}{m^2_g}\bigg)\bigg[\frac{1}{z_+}- \ln\frac{\nu}{\kappa} \delta(\ell)\bigg]\bigg\}\,.
\eea
The single logarithm of $\nu$ in this expression is minimized for $\nu \sim \kappa \sim Q(1-x)$, which is different from the value of $\nu\sim Q$ required to minimize the logarithm in Eq.~(\ref{finalcoll}). Thus, while the dependence on the rapidity regulator vanishes if the above expression is added to the collinear result in Eq.~(\ref{finalcoll}), a single large logarithm of the ratio of $Q$ to $\kappa\sim Q(1-x)$ is left over. This constitutes an incomplete cancelation of sensitivity to rapidity scales between the soft and collinear contributions, and running in $\nu$ is necessary to resum these logarithms.

\section{Renormalization \& Running}
\label{rnr}

The divergences in $\epsilon$ and $\eta$ in Eq.~(\ref{finalcoll}) and Eq.~(\ref{finalsoft}) can be subtracted by suitable counter terms, which we define by
\bea
{\cal C}_{n}( Q-k)^{R} &=&Z_{n}^{-1} {\cal C}_{n}( Q-k)^{B} \nn\\
S(\ell)^{R} &=& \int d\ell' Z_s(\ell-\ell')^{-1}S(\ell')^{B} \,, \nn
\eea
where the superscripts $R$ and $B$ indicate renormalized and bare. To extract $Z_{n}$ we need the wave function renormalization factor at one loop
\beq
Z_\psi = 1 - \frac{\alpha_s C_F}{4 \pi \, \epsilon} \,.
\eeq
Then the one loop collinear counter term is
\beq
\label{collct}
Z_{n} = 1 + \frac{ \alpha_s C_F}{\pi}w^2\bigg[\frac{e^{\epsilon \gamma_E} \Gamma(\epsilon) }{ \eta}\bigg(\frac{\mu}{m_g}\bigg)^{2 \epsilon}+
\frac{1}{\epsilon}\bigg( \frac{3}{4} + \ln \frac{\nu}{\bn\cdot p}\bigg)\bigg]\,.
\eeq
The one-loop soft counter term is 
\beq
\label{softct}
Z_s(\ell) = 
\delta(\ell) + \frac{\alpha_s C_F}{\pi}w^2\bigg\{- \frac{e^{\epsilon \gamma_E} \Gamma(\epsilon) }{ \eta}\bigg(\frac{\mu}{m_g}\bigg)^{2 \epsilon}\delta(\ell)
+\frac{1}{\epsilon} \bigg[\frac{1}{(\ell)_+} -\ln{\nu}\delta(\ell)\bigg]\bigg\}\,.
\eeq

A non-trivial check on this result is to verify that these counter terms obey the consistency condition
\beq
Z_H Z_{J_{\bn}}(\ell) = Z_{n}^{-1} Z_s^{-1}(\ell)\,,
\eeq
where $Z_H$ is the square of the counter term of the SCET DIS current, and $Z_{J_{\bn}}(\ell)$ is the jet-function counter term. The one loop expression for $Z_H$ was first given in the appendix of Ref.~\cite{Bauer:2003di}. Converting their expression from $4 -\epsilon$ dimensions to $4 - 2 \epsilon$ dimensions and squaring gives
\beq
Z_H = 1 - \frac{\alpha_s C_F}{2\pi} \bigg( \frac{2}{\epsilon^2}+ \frac{3}{\epsilon}+ \frac{2}{\epsilon}\ln \frac{\mu^2}{Q^2}\bigg)\,,
\eeq
where $Q^2 = \bn \cdot p\, n\cdot p_X$, with $p^\mu_X$ the final state momentum.
The one-loop expression for $Z_{J_{\bn}}(\ell)$ can be obtained from Ref.~\cite{Manohar:2003vb}
\beq
Z_{J_{\bn}}(\ell) = \delta(\ell) +  \frac{\alpha_s C_F}{4\pi}\bigg[ \bigg(\frac{4}{\epsilon^2}+\frac{3}{\epsilon}-\ln\frac{n\cdot p_X}{\mu^2}\bigg)\delta(\ell) -\frac{4}{\epsilon}\frac{1}{\ell_+}\bigg] \,.
\eeq
Thus at one loop
\beq
Z_H Z_{J_{\bn}}(\ell)=   
\delta(\ell) +  \frac{\alpha_s C_F}{4\pi}\bigg\{ \bigg[-\frac{3}{\epsilon}+\frac{4}{\epsilon}\ln (\bn\cdot p)\bigg]\delta(\ell) -\frac{4}{\epsilon}\frac{1}{\ell_+}\bigg\} \,.
\eeq
Adding the inverse of Eq.~(\ref{collct}) and the inverse of Eq.~(\ref{softct}) we find that at one loop  the expression for $Z_{n}^{-1} Z_s^{-1}(\ell)$ agrees with the above expression  satisfying the consistency condition.

We can extract the one-loop anomalous dimensions from the counter terms above. The $\mu$ anomalous dimensions are
\bea
\label{muanondim}
\gamma^\mu_{n} (\mu,\nu)&=& \frac{2 \alpha_s C_F}{\pi}\bigg(\frac{3}{4}+\ln\frac{\nu}{\bn\cdot  p_p}\bigg)\\
\gamma^\mu_s(\ell;\mu,\nu) &=& \frac{2 \alpha_s C_F}{ \pi}\bigg[\frac{1}{\ell_+} -\ln \nu \, \delta(\ell)\bigg] \,.\nn
\eea
When $\gamma^{\mu}_{n}$ and $\gamma^{\mu}_{s}$ are added the rapidity scale $\nu$ cancels as it must, however we clearly see that a large logarithm of $\bn\cdot p \sim Q$ remains. Once again this is a manifestation of the incomplete cancelation of rapidity logarithms.
The $\nu$ anomalous dimensions are
\bea
\label{nuanondim}
\gamma^\nu_{n} (\mu,\nu)&=& \frac{\alpha_s C_F}{\pi}\ln\frac{\mu^2}{m_g^2}\\
\gamma^\nu_s(\mu,\nu) &=& - \frac{\alpha_s C_F}{\pi}\ln\frac{\mu^2}{m_g^2}\,.\nn
\eea
We notice that $\gamma^{\nu}_{n} + \gamma^{\nu}_{s}=0$ as required for consistency, and 
as is immediately obvious from the presence of $m_g$ in these expressions, the running in $\nu$ is {\it not} perturbative. Although we have calculated these anomalous dimensions in perturbation theory at one loop in a particular scheme, they reveal sensitivity to IR scales, which may signal a breakdown of rapidity factorization in $\SCETb$. This IR sensitivity is also present if a $\delta$ regulator~\cite{Chiu:2009yx} is used to regulate rapidity divergences~\cite{fz}

The running in $\mu$ and the running in $\nu$ are independent of each other and can be carried out in any order. 
The one-loop $\mu$-running factor for the collinear function is
\bea
{\cal C}_{n}(k;\mu,\nu_c) &=& U(\mu,\mu_0,\nu_c){\cal C}_{n}(k;\mu_0,\nu_c) \\
U(\mu,\mu_0,\nu_c)&=&e^{\frac{3}{4}\omega(\mu,\mu_0)} \bigg[\frac{\nu_c}{\bn\cdot p}\bigg]^{\omega(\mu,\mu_0)}
\,, \nn
\eea
where $\nu_c$ is the collinear rapidity scale and
\beq
\omega(\mu,\mu_0) = \frac{4 C_F}{\beta_0}\ln\bigg[\frac{\alpha_s(\mu)}{\alpha_s(\mu_0)}\bigg]\,.
\eeq
The one-loop $\mu$-running factor for the soft function is
\bea
S(\ell;\mu,\nu_s) &=& \int d r \, U(\ell-r;\mu,\mu_0,\nu_s) S(r;\mu_0,\nu_s)\\
 U(\ell-r;\mu,\mu_0,\nu_s)&=&\frac{ \big( e^{2 \gamma_E}\nu_s\big)^{-\omega(\mu,\mu_0)}}{\Gamma(\omega(\mu,\mu_0))}\frac{1}{[(\ell-r)^{1- \omega(\mu,\mu_0)}]_+}\,. \nn
\eea
Despite the fact that the $\nu$ running is non-perturbative we give the expression for $\nu$ running factor
\bea
\label{softnurun}
S(\ell;\mu_s,\nu)&=& V(\mu_s, \nu,\nu_0) S(\ell;\mu_s,\nu_0)\\
V(\mu_s, \nu,\nu_0)&=& \bigg[\frac{\nu}{\nu_0}\bigg]^{\omega(\mu_s,m_g)} \,.\nn
\eea
Since the running in $\mu$ is independent of the running $\nu$ we are free to choose the order in which we resum the different types of logarithms. Here, however, the running in $\nu$ is non-perturbative and can not be done. Thus we can only run in $\mu$. One approach is to carry out the running in $\mu$ with $\nu_c\sim Q$ and $\nu_s \sim \Lambda_{\rm{had}}$ in the expressions above. This minimizes logarithms of $\nu$ however the $\nu$ dependence of the soft and collinear pieces does not cancel in the perturbative expressions for the the one loop results nor does it cancel between the anomalous dimensions. If the $\nu$ running were not IR sensitive and could be carried out there would be an additional resumation factor that would result from he running the soft function from $\nu_{s}$ to $\nu_{c}$. Once this factor is included all expressions would be $\nu$ independent to the order we are working.  A second approach, equivalent to the one adopted in Ref.~\cite{Becher:2006mr}, is where the $\nu$ scale in the one-loop matrix elements and the anomalous dimensions are all chosen to be the same. In this approach the $\nu$ dependence cancels between the soft and collinear pieces, but a large single logarithm is left over. Implicitly this approach first runs the soft function in $\nu$ to the scale $\nu_s \sim \nu_{c}\sim Q$, where $\mu \sim \LQCD$ in Eq.~(\ref{softnurun}). Since the running in $\nu$ is non-perturbative we are left with little choice but to include the $\nu$ resumation factor as part of our nonperturbative model for the PDF. As a result the model could contain large single logarithms that would manifest themselves as larger than expected parameters.

\section{Definition of the parton distribution function}
\label{pdf}
Finally we consider the definition of the parton distribution function. The PDF defined in Eq.~(\ref{phi}) above is worrisome because the soft function is sensitive to both the initial and final state (due to the soft Wilson lines running to infinity). This would imply that the PDF is not universal to other process with the same initial state but different final state. 
To keep the PDF universal, we want to require that the PDF only depend on properties of the initial state. 
In this section we show that the soft function in Eq.~(\ref{softfun}) can be manipulated into a form which is only sensitive to initial state radiation making our definition of the PDF universal. 

We introduce Wilson lines linking the far past to the far future~\cite{Arnesen:2005nk}
\bea
\tilde Y^{\infty\dag}_\bn=\bar{P}\exp\left(-ig\int_{-\infty}^{\infty}ds \bn\cdot A_s(\bn s)\right) \\
\tilde Y^{\infty}_\bn= P\exp\left(ig\int_{-\infty}^{\infty}ds \bn\cdot A_s(\bn s)\right)
\eea
and insert the identity $\tilde Y^{\infty\dag}_\bn \tilde Y^{\infty}_\bn\equiv 1$ between the time-ordered and anti-time-ordered Wilson lines in the soft function \req{softfun}.  In Appendix \ref{app:wilsonlines}, we show that
\bea
&\frac{1}{N_c}&\langle 0|\textrm{Tr}\left(\bar{\textrm{T}}\left[Y^\dag_n(n\cdot x)\tilde Y^\dag_\bn(n\cdot x)\right]\tilde Y^{\infty\dag}_\bn\tilde Y^{\infty}_\bn\textrm{T}\left[\tilde Y_\bn(0)Y_n(0)\right]\right)|0\rangle \nn \\
& &=  \frac{1}{N_c}\langle 0|\textrm{Tr}\left(\bar{\textrm{T}}\left[Y^\dag_n(n\cdot x)Y_\bn(n\cdot x)\right]\textrm{T}\left[Y^\dag_\bn(0)Y_n(0)\right]\right)|0\rangle 
\equiv  \int d\ell e^{\frac{i}{2}ln\cdot x}S(\ell,\mu) \label{initY}
\eea
which gives an $S(\ell,\mu)$ that is sensitive only to initial state information, since all four Wilson lines extend from minus infinity to the interaction point. Now the expression for the PDF defined in Eq.~(\ref{phi}) has the form
\bea\label{newPDF}
\phi_q^{ns}(z;\mu) &=&
\frac{1}{2} \sum_\sigma  
 \langle{h_{n}(p,\sigma)}| \bar{\chi}_{n}(0)\frac{\bnslash}{2}   \chi_{n}(0) |{h_{n}(p,\sigma)}\rangle\\
 && \times \int \frac{d n\cdot x}{4 \pi} e^{\frac{i}{2}Q z n\cdot x}
 \frac{1}{N_c} \langle{0} |\textrm{Tr}\bigg(
\bar{\textrm{T}}\bigg[ Y^\dagger_{n}(n\cdot x) {Y}_{\bn}(n\cdot x) \bigg]
\textrm{T}\bigg[{Y}^\dagger_{\bn}(0) Y_{n}(0) \bigg]
\bigg)|0\rangle\,, \nn
\eea
which makes the it manifest that the PDF only depends on the initial state. 

\section{Conclusions}

In this paper we have revisited DIS in the endpoint region $x \sim 1$ with the goal of a clearer understanding of the individual factors in the factorized hadronic tensor. We use a two-step process where we first match QCD onto \SCETa at a scale $\sim Q$ and then match onto \SCETb at a scale $\sim Q\sqrt{1-x}$. In agreement with previous results we find that the hadronic tensor factors into the form
\begin{equation*}
W_{\rm eff}^{\mu\nu}=-g_{\perp}^{\mu\nu}H(Q;\mu_q,\mu_c)\int d\ell J_{\bn}(\ell;\mu_c,\mu))\phi_q^{ns}(Q\frac{1-x}{x}+\ell;\mu)
\end{equation*}
with $H$ the hard coefficient, $J_{\bar n}$ the jet function, and $\phi^{ns}_q$ the quark PDF.  The PDF is defined as 
\begin{equation*}
\phi^{ns}_q(\ell;\mu)=\mathcal{Z}_n(Q;\mu,\nu)S(\ell;\mu,\nu)
\end{equation*}
with $\mathcal{Z}_n$ the collinear factor and $S$ the soft function.  Both the collinear factor and soft function need a rapidity regulator to be well defined, while the product is free of rapidity divergences.  However, as we show in a one-loop calculation, the scale which minimizes rapidity logarithms in the collinear factor is $\nu_c\sim Q$, while the scale which minimizes rapidity logarithms in the soft function is $\nu_s\sim Q(1-x)$.  Thus, while the product of $\mathcal{Z}_n$ and $S$ is free of rapidity divergences, there is only an incomplete cancelation of these divergences, which results in a $\ln(\nu_s/\nu_c)$ term in the PDF.  To sum this large logarithm, running rapidity is necessary.  We find that rapidity running in DIS at the endpoint is nonperturbative and has to be absorbed into the nonperturbative soft function.

In addition, we find some interesting aspects to the one loop calculations. First, in the collinear factor, real radiation is prohibited by label momentum conservation so this function only includes virtual contributions. Second, in the one loop computation of the soft function, the overlap of the soft degrees of freedom with $n$ and $\bn$ collinear degrees of freedom needs to be subtracted. 

Finally, we consider the proper definition of the PDF. Our derivation of the factored form of the DIS hadronic tensor makes explicit that while the collinear factor only depends on the initial state interactions, the soft function appears to depend both on initial and final state interactions. We show that appearances can be deceiving and that the soft function can be manipulated into a form that is sensitive only to initial state information which guarantees the universality of the PDF. In a future publication we will examine rapidity divergences in different regularization schemes both in DIS and Drell-Yan in the endpoint region~\cite{fz}.

\acknowledgments
We would like to thank Aneesh Manohar and Wouter Waalewijn for discussions and helpful comments.
This work was supported in part by the Director, Office of Science, Office of Nuclear Physics, of the U.S. Department of Energy under grant numbers DE-FG02-06ER41449 and DE-FG02-04ER41338. S.F.  also acknowledges support from the DFG cluster of excellence ``Origin and structure of theuniverse''.

%%%%

\begin{appendix}

\section{Initial and Final State Soft Wilson Lines in Soft Functions}\label{app:wilsonlines}

In this appendix we prove \req{initY}, based on the work in the appendix of Ref.~\cite{Bauer:2003di}.  We start with a general event-shape function,
\begin{align}\label{pfstep1}
S(k) &= \frac{1}{N_c}\int\frac{du}{(2\pi)}e^{iku}\langle 0| \bar T[(Y_{\bar n}^{\dagger})_d^{e}(Y_n)_e^a](un/2)T[(Y_{n}^{\dagger})^c_a(Y_{\bar n})_c^d](0)|0\rangle
\end{align}
The Wilson lines in this expression can be divided into $N$ infinitesimal segments of length $ds$ with a subscript denoting their space-time position along the integration path,
\begin{align}
\label{Yndefn} (Y_{n})_e^a& = \overline{P}\exp(-ig\int_0^\infty ds n\cdot A_s) = (e^{-igA_1ds})_e^{b_1}\ldots(e^{-igA_Nds})_{b_{N-1}}^a \,, \\
\label{Yndagdefn} (Y_{n}^{\dagger})_a^c &= {P}\exp(ig\int_0^\infty ds n\cdot A_s) = (e^{igA_Nds})_a^{b_{N-1}}\ldots(e^{igA_1ds})_{b_1}^c \,, \\
(Y_{\bar n})_c^d &= {P}\exp(ig\int_{-\infty}^0 ds \bar n\cdot A_s) = (e^{-igA_1(\bar n)ds})_a^{b_1}\ldots(e^{-igA_N(\bar n)ds})_{b_{N-1}}^c \,, \\
(Y_{\bar n}^\dagger)_d^{e} &= \overline{P}\exp(ig\int_{-\infty}^0 ds \bar n\cdot A_s) = (e^{igA_N(\bar n)ds})_d^{b_{N-1}} \ldots (e^{igA_1(\bar n)ds})_{b_1}^e \,.
\end{align}
Among these Wilson lines, \req{Yndefn} and \req{Yndagdefn} are sums of outgoing gluons, which represent final state gluons.  Applying time-ordering and anti-time-ordering operators, we obtain
\begin{align}
T(Y_{n}^{\dagger})_a^c&=(Y_{n}^{\dagger})_a^c \,,
\end{align}
and
\begin{align}
\bar T(Y_{n})_e^a&=(Y_{n})_e^a \,.
\end{align}
For the other two we find
\begin{align}
T(Y_{\bar n})_c^d &= (e^{-igA_N(\bar n)ds})_{b_{N-1}}^d\ldots(e^{-igA_1(\bar n)ds})_c^{b_1}\\
&=(e^{-igA_N^T(\bar n)ds})_d^{b_{N-1}}\ldots(e^{-igA_1^T(\bar n)ds})_{b_1}^c\\
&=(e^{ig\bar n\cdot \bar A_N ds})_d^{b_{N-1}}\ldots(e^{ig\bar n\cdot A_1ds})_{b_1}^c = (\bar{Y}_{\bar n}^{\dagger})_d^c\,, \\
%%%%%%%%%
\bar T(Y_{\bar n}^\dagger)_d^{e} &= (e^{igA_1(\bar n)ds})_{b_1}^e\ldots(e^{igA_N(\bar n)ds})_d^{b_{N-1}}\\
&=(e^{igA_1^Tds})_e^{b_1}\ldots(e^{igA_N^T(\bar n)ds})_{b_{N-1}}^d\\
&=(e^{-ig\bar A_i\bar nds})_e^{b_1}\ldots(e^{-ig\bar n\cdot \bar A_Nds})_{b_{N-1}}^d=\bar Y_{\bar ne}^d \,.
\end{align}
Applying the above identities to the expression in Eq.~(\ref{pfstep1}) gives
\begin{align}
S(k)&=\frac{1}{N_c}\int\frac{du}{(2\pi)}e^{iku}\langle 0| (\overline{Y}_{\bar n}^{\dagger})_e^d(Y_{n})_e^{a'}(un/2)\delta^a_{a'} (Y_{n}^\dagger)_a^{c}(\overline{Y}_{\bar n})_d^c(0)|0\rangle \,.
\end{align}
Now consider two infinite Wilson lines
\begin{align}
(Y_\infty)_{a'}^f&=P\exp\{ig\int_{-\infty}^\infty dsn\cdot A_s(\frac{un}{2})\}_{a'}^f=P\exp\{ig\int_{-\infty}^\infty dsn\cdot A_s(0)\}_{a'}^f\\
&=\{(e^{igA_N\cdot nds})_{a'}^{c_{N-1}}\cdot (e^{igA_1\cdot nds})_{c_1}^{c_0}\}\{(e^{igA_{-1}\cdot nds})_{c_0}^{c_1}\cdot (e^{igA_{-N}\cdot nds})_{c_{N+1}}^f\}\,, \\
(Y_\infty^\dagger)_f^a &= \overline{P}\exp\left(-ig\int_{-\infty}^{\infty}ds n\cdot A_s(un/2)\right)^{a'}_f=\overline{P}\exp\left(-ig\int_{-\infty}^{\infty}ds n\cdot A_s(0)\right)^{a'}_f  \\
&= \{(e^{-igA_{-N}\cdot nds})_f^{c_{N+1}}\cdot (e^{-igA_{-1}\cdot nds})_{c_{-1}}^{c_0}\}\{(e^{-igA_1\cdot nds})_{c_0}^{c_1}\cdot (e^{-igA_N\cdot nds})_{c_{N-1}}^a\} \,,
\end{align}
which have the property that 
\begin{align}
(Y_\infty)_{a'}^f(Y_\infty^\dagger)_f^a &=\delta_{a'}^a \,.
\end{align}
We can use this property to replace the identity $\delta_{a'}^a$ in $S(k)$ with the pair of infinite Wilson lines above
\begin{align}
S(k)=&\frac{1}{N_c}\int\frac{du}{(2\pi)}e^{iku}  \langle 0|(\bar Y_{\bar n})_e^d (Y_{n})_e^{a'}(un/2)\delta_{a'}^a (Y_{n}^{\dagger})_a^c(\bar Y_{\bar n}^{\dagger})_d^c(0)|0\rangle\\
=&\frac{1}{N_c}\int\frac{du}{(2\pi)}e^{iku}  \times \notag \\
&\langle 0|\{(e^{-ig\bar A_N\cdot\bar nds})_e^{b_1}\ldots(e^{-ig\bar n\cdot A_1 ds})_{b_{N-1}}^d\}\{(e^{-igA_1\cdot nds})_e^{b_1}\ldots(e^{-igA_N\cdot nds})_{b_{N-1}}^{a'}\}(\frac{un}{2}) \notag \\
&\cdot\{(e^{igA_N\cdot nds})_{a'}^{c_{N-1}}\ldots(e^{igA_1\cdot nds})_{c_1}^{c_0}\}\{(e^{igA_{-1}\cdot nds})_{c_0}^{c_{-1}}\ldots (e^{igA_{-N}\cdot nds})_{c_{N+1}}^f\}(\frac{un}{2})  \notag \\
&\cdot\{(e^{-igA_{-N}\cdot nds})_f^{c_{-N+1}}\ldots(e^{-igA_{-1}\cdot nds})_{c_{-1}}^{c_0}\}\{(e^{-igA_1\cdot nds})_{c_0}^{c_1}\ldots(e^{-igA_N\cdot nds})_{c_{N-1}}^a\}(0)   \notag \\
&\cdot\{(e^{igA_N\cdot n ds})_a^{b_{N-1}}\ldots (e^{igA_1\cdot nds})_{b_1}^c\}\cdot\{(e^{ig\bar n\cdot A_N ds})_d^{b_{N-1}}\ldots(e^{ig\bar n\cdot A_1ds})_{b_1}^c\}(0)  |0\rangle   \notag \\
=&\frac{1}{N_c}\int\frac{du}{(2\pi)}e^{iku}  \times \notag \\
&\langle 0| \{(e^{-ig\bar A_N\cdot\bar n ds})_e^{b_1}\ldots(e^{-ig\bar n\cdot A_1ds})_{b_{N-1}}^d\}\{(e^{igA_{-1}\cdot nds})_e^{c_{-1}}\ldots(e^{igA_{-N}\cdot nds})_{c_{-N+1}}^f\}(\frac{un}{2})  \notag \\
&\cdot\{(e^{-igA_{-N}\cdot nds})_f^{c_{-N+1}}\ldots(e^{-igA_{-1}\cdot nds})_{c_{-1}}^c\}\{(e^{ig\bar n\cdot A_Nds})_d^{b_{N-1}}\ldots(e^{ig\bar A_1\cdot\bar nds})_{b_1}^c\}(0) |0\rangle \notag \\
\label{pffinal}
=& \frac{1}{N_c}\int\frac{du}{(2\pi)}e^{iku}  \langle 0| (\overline{Y}_{\bar n})_e^d(Y_n)_e^f(un/2)(Y^\dagger_n)^f_c(Y^\dagger_{\bar n})_d^c(0) |0\rangle
\end{align}
in which
\begin{align}
(Y_n)^f_e(un/2) &= \left( e^{igA_{-1}\cdot nds}\right)^{c_{-1}}_e\ldots \left(e^{igA_{-N}\cdot nds}\right)^f_{c_{-N+1}} \\
 &= P \exp\left( ig\int_{-\infty}^0ds n\cdot A_s\right) \notag \\
(Y^\dagger_n)_c^f(0) &=   \left( e^{-igA_{-N}\cdot nds}\right)^{c_{-N+1}}_f\ldots \left(e^{-igA_{-1}\cdot nds}\right)^c_{c_{-1}} \\
 &= \overline{P} \exp\left( -ig\int_{-\infty}^0ds n\cdot A_s\right) \notag
\end{align}
are incoming gluons.  Thus, from \req{pfstep1} to \req{pffinal}, we show that by inserting the identity operator for infinite Wilson lines, we change the final state Wilson lines in the soft function into initial state Wilson lines.

\end{appendix}

\bibliography{DISEndpoint}

\end{document}